\documentclass[11pt,titlepage]{article}
\renewcommand{\arraystretch}{1.4}
\input amssym.def
\input amssym.tex

\font\sevenrm=cmr7

\font\ninerm=cmr9
\font\tenrm=cmr10

\font\Bbb=msbm10
\input macros.pic
\usepackage[normal]{caption}
\usepackage{enumerate}

\begin{document}

\pagestyle{plain}
\pagenumbering{arabic}
\def\thesection{\Roman{section}.}
\def\thesubsection{\Roman{section}-\Alph{subsection}.}
\def\thesubsubsection
  {\Roman{section}-\Alph{subsection}-\arabic{subsubsection}.}
\def\theequation{\Roman{section}.\arabic{equation}}
\def\<{\begin{equation}}
\def\>{\end{equation}}
\def\rf#1{(\ref{#1})}
\def\taub{\overline{\tau}}
\def\rhob{\overline{\rho}}
\def\prodd{\displaystyle\prod}
\def\sumd{\displaystyle\sum}
\def\C{{\cal C}}
\def\I{{\cal I}}
\def\J{{\cal J}}
\def\K{{\cal K}}
\def\R{\mbox{\Bbb R}}
\def\S{{\cal S}}
\def\qp{\overrightarrow{\kern -.1em q\kern .1em},
  \overrightarrow{\kern -.13em p\kern .13em}}
\def\strutdown{\vtop to 3.7pt{}}
\def\strutup{\vbox to 9.2pt{}}
\def\rectbox#1#2{\beginpicture \frame <1.5pt> {\makebox[#1in]{\ninerm
   \vphantom{$'$}#2}} \endpicture}
\def\rectfbox#1{\fbox{\vphantom{t}#1}}
\def\rndbox#1#2{\beginpicture \setcoordinatesystem units <.1in,.1in>
   \put {\oval(#1,11.2)} at 0 0.05 \put {\vphantom{$'$}\ninerm #2} at 0 .45 \endpicture}

\newtheorem{lemma}{Lemma}
\renewcommand{\thefootnote}{\fnsymbol{footnote}}
\title{
\hfill CENS/T04/004 \\
\vfill
\bf Marginal distributions in ($\bf 2N$)-dimensional
phase space and the quantum ($\bf N+1$) marginal theorem\thanks{Work
supported by the Indo-French Centre for the Promotion of Advanced
Research, Project Nb 1501-02.}.}
\author{{\Large G. Auberson}\thanks{e-mail: auberson@lpm.univ-monpt2.fr} \\
	\sl Laboratoire de Physique Math\'ematique, UMR 5825-CNRS, \\
	\sl Universit\'e Montpellier II, \\
	\sl F-34095 Montpellier, Cedex 05, FRANCE.
\and
	{\Large G. Mahoux}\thanks{e-mail: mahoux@spht.saclay.cea.fr} \\
	\sl Service de Physique Th\'eorique, \\
	\sl Centre d'\'Etudes Nucl\'eaires de Saclay, \\
	\sl F-91191 Gif-sur-Yvette Cedex, FRANCE.
\and
	{\Large S.M. Roy}\thanks{e-mail: shasanka@theory.tifr.res.in} and
	{\Large Virendra Singh}\thanks{e-mail: vsingh@theory.tifr.res.in} \\
	\sl Department of Theoretical Physics, \\
	\sl Tata Institute of Fundamental Research, \\
	\sl Homi Bhabha Road, Mumbai 400 005, INDIA.
}
\date{February 14, 2004.}
\maketitle
\renewcommand{\thefootnote}{\arabic{footnote}}

\begin{abstract}
We study the problem of constructing a probability density in
$2N$-dimensional phase space which reproduces a given collection of
$n$ joint probability distributions as marginals. Only distributions
authorized by quantum mechanics, i.e. depending on a (complete)
commuting set of $N$ variables, are considered. A diagrammatic or
graph theoretic formulation of the problem is developed.  We then exactly
determine the set of ``admissible'' data, i.e. those types of
data for which the problem always admits solutions. This is done in
the case where the joint distributions originate from quantum
mechanics as well as in the case where this constraint is not
imposed. In particular, it is shown that a necessary (but not
sufficient) condition for the existence of solutions is $n\leq
N+1$. When the data are admissible and the quantum constraint is not
imposed, the general solution for the phase space density is
determined explicitly. For admissible data of a quantum origin, the
general solution is given in certain (but not all) cases. In the
remaining cases, only a subset of solutions is obtained.
\end{abstract}

\section{Introduction}
\setcounter{equation}{0}

In quantum theory, we usually assume that probability densities for
eigenvalues of two noncommuting observables cannot be measured by the
same experimental set up. A related theoretical question is : does
there exist a joint probability distribution of the eigenvalues of two
such observables $A$ and $B$ which correctly reproduces the individual
probabilities for $A$ and for $B$ as marginals (i.e., on integration
over the eigenvalues of the other observable). Perhaps surprisingly,
the answer to this question is yes.  More generally, for a system
with $N$ configuration space variables $q_1,q_2,\ldots,q_N$, consider
complete commuting sets (CCS) $S_1,S_2,\ldots,S_n$ of observables,
each $S_i$ consisting of some coordinate and some momentum variables
(where $S_i$ and $S_j$ must contain some mutually non-commuting
observables to be considered distinct CCS).  Is there a joint
probability density $\rho(q_1,\ldots,q_N,p_1,\ldots,p_N)$ whose
marginals reproduce the quantum probability densities of the different
CCS, $S_1,S_2,\ldots,S_N$? We shall prove that a necessary condition
for this to be possible for arbitrary quantum states is $n\leq N+1$;
this result is a precise no go theorem for simultaneous realization of
more than $(N+1)$ quantum marginals.

Actually, this no go theorem also has applications to the classical
arena of joint time-frequency distributions in signal processing and
joint position-wave number distributions in image processing. It is
therefore useful to state the problem in a general setting
encompassing both classical and quantum mechanics.

Consider the general problem of reconstructing a probability density
over $\R^M$, given a set of $n$ associated joint probability
distributions over subspaces of $\R^M$. In this general setting the
problem can be stated as follows. Suppose first that a probability
density $\rho(y_1,\ldots,y_M)$ is given and define the marginal
distributions
\< \sigma_\alpha(Y_\alpha) = \int dY_\alpha'\,
  \rho(y_1,\ldots,y_M) \qquad\qquad (\alpha=1,\ldots,n)\,, \label{I.1}
\>
where $Y_\alpha\bigcup\,Y_\alpha'$ is, for each $\alpha$, a partition
of $\{y_1,\ldots,y_M\}$. These joint distributions obey a set of
compatibility conditions. Indeed, let $Y_{\alpha\beta}$ be the set of
variables that $\sigma_\alpha(Y_\alpha)$ and $\sigma_\beta(Y_\beta)$
have in common and introduce the partitions
$Y_\alpha=Y_{\alpha\beta}\bigcup Y_{\alpha\beta}'$ and
$Y_\beta=Y_{\alpha\beta}\bigcup Y_{\alpha\beta}''$. Then eqs.\,\rf{I.1}
imply
\< \int dY_{\alpha\beta}'\,\sigma_\alpha(Y_{\alpha\beta},Y_{\alpha\beta}')
  =\int dY_{\alpha\beta}''\,\sigma_\beta(Y_{\alpha\beta},Y_{\alpha\beta}'')
  \,. \label{I.2}
\>
Conversely, suppose that a set $\{\sigma_1,\ldots,\sigma_n\}$ of joint
probability distributions is given, which satisfies the compatibility
conditions \rf{I.2}. Is it always possible to find some probability
density $\rho$ which reproduces them as marginals, in accordance with
equations \rf{I.1}? In the affirmative, how can we ``reconstruct'' such
$\rho$'s? It turns out that eqs.\,\rf{I.2} are in general only necessary
conditions for the existence of a positive density $\rho$, and our problem is
precisely to solve the questions of existence, multiplicity and
explicit determination of the $\rho$'s (if any).

Actually, we address these questions not in such a general setting,
but in the case where $\R^M=\R^{2N}$ is the phase space of some
physical system with coordinates $y_j$ identified with conjugate
canonical variables $\{q_j,p_j\}_{j=1}^N$ and where all distributions
$\sigma_\alpha$ depend on exactly $N$ variables $x_1,\ldots,x_N$
restricted by the condition $x_i=q_i$ or $p_i$ ($i=1,\ldots,N$). The
reason for this choice lies in the motivation of the problem within
the context of quantum mechanics. Indeed, the functions $\sigma_\alpha$
just introduced can then be understood as quantum probability
distributions associated with $n$ complete commuting sets of
observables (CCS), selected among $2^N$ possible choices (the $2^N$
possible assignments of the variables $x_i$). By ``{\sl quantum}
probability distributions'' is meant here a set of functions
$\sigma_\alpha(x_1,\ldots,x_N)$ derived from a common wave function
$\langle q_1,\ldots,q_N|\psi\rangle$ (in the Schr\"odinger
representation), in accordance with the formula
\< \sigma_\alpha(x_1,\ldots,x_N)=|\langle x_1,\ldots,x_N|\psi\rangle|^2
   \qquad (\alpha=1,\ldots,n)\,, \label{I.3}
\>
or more generally, for a mixed quantum state described by the density
operator $\hat{\rho}$,
\< \sigma_\alpha(x_1,\ldots,x_N) =
   \langle x_1,\ldots,x_N|\,\hat{\rho}\,|x_1,\ldots,x_N\rangle
   \qquad (\alpha=1,\ldots,n)\,. \label{I.4}
\>

The problem in this physical framework is directly related to the
construction of ``maximally realistic quantum mechanics'', a program
initiated by S.M.~Roy and V.~Singh in 1995 \cite{RS1} and intensively
pursued since then \cite{RS2}-\cite{AMRS4}. Without entering a
detailed discussion of this relationship from the viewpoint of quantum
physics (for which we refer the reader to \cite{RS1}-\cite{AMRS4}),
let us recall the main results gained so far. In
\cite{RS2} it was shown that for any $N\geq2$, a set of $n=N+1$
quantum probability distributions of the special form
\< \{\sigma_1(q_1,q_2,\ldots,q_N),\sigma_2(p_1,q_2,\ldots,q_N),
  \sigma_3(p_1,p_2,q_3,\ldots,q_N),\ldots,\sigma_{N+1}
  (p_1,p_2,\ldots,p_N)\} \label{I.5}
\>
can be realized as a set of marginals of a common phase space
probability density
$\rho(q_1,\ldots,q_N,$ $p_1,\ldots,p_N)$. Further the ``no go''
conjecture was made that
for $n\geq N+2$ (and for any choice of $n$ distinct CCS), there exist
sets $\{\sigma_1,\ldots,\sigma_n\}$ of quantum probability
distributions which cannot be recovered as marginals of some $\rho$. The
determination of the most general density $\rho$ reproducing the set
\rf{I.5}, as well as the status of sets of CCS-distributions different
from \rf{I.5} and not necessarily of a quantum origin (i.e. mutually
compatible but not necessarily construed according to eqs.\,\rf{I.4}),
were left as open questions. In \cite{AMRS4}, hereafter denoted by
(I), complete answers to these questions were given in the special
case $N=2$ (see also \cite{AMRSl} for a brief summary). In particular,
the general positive solution $\rho$ of the equations
\< \left\{ \begin{array}{rcl}
 \displaystyle\int dp_1dp_2\,\rho(q_1,q_2,p_1,p_2)&=&\sigma_1(q_1,q_2)\,,\\
 \displaystyle\int dq_1dp_2\,\rho(q_1,q_2,p_1,p_2)&=&\sigma_2(p_1,q_2)\,,\\
 \displaystyle\int dq_1dq_2\,\rho(q_1,q_2,p_1,p_2)&=&\sigma_3(p_1,p_2)\,,
 \end{array} \right. \label{I.6}
\>
was worked out for an arbitrary set $\{\sigma_1,\sigma_2,\sigma_3\}$
(quantum or not) and the ``no go'' conjecture stated above was proved
for $N=2$. In fact, it was shown that a necessary and sufficient
condition for $n$ ($\leq4$) arbitrarily given compatible
$\sigma_\alpha$'s to be marginals of a probability density in
4-dimensional phase space is $n\leq3$ (``Three marginal
theorem''). These results were obtained by first deriving certain
correlation inequalities between the $\sigma_\alpha$'s from the mere
existence of a positive and normalized $\rho$. Such inequalities,
which are the analogues in phase space of the standard Bell
inequalities for spin variables \cite{B}, turn out to have an interest
of their own in the context of quantum physics, as discussed in
\cite{AMRSl}-\cite{AMRS4}.

The generalization of the study performed in (I) to the case of an
arbitrary number $n$ of CCS-distributions of any species in a phase
space of arbitrary dimension $2N$, which is precisely the aim of the
present work, is not a straightforward task. It will be accomplished
by means of two main tools : the Bell-like inequalities just mentioned
(it turns out that no new correlation inequalities, proper to the
$2N$-dimensional case, are needed for the present purpose) and a
specific diagrammatic formulation of the problem which appears 
essential both for a concise exposition of our final statements and
for their proof. In this way, we shall be able to treat the problem
exhaustively and to give, in the general case, definite answers to
the questions previously posed, in the form of a clear-cut
theorem. This theorem will be stated at once in section II
(Theorem\,\ref{theorem1}), after having introduced a set of
appropriate definitions. As a by-product, the theorem affords a proof
of the ``no go'' conjecture for any $N$ (Theorem 2).  On the positive side,
it considerably extends early results of Cohen and
Zaparovanny \cite{CZ} for two marginals with non
intersecting sets of variables by simultaneous realizability of $N+1$
marginals which have intersecting sets of variables as well. The rest
of the paper (sections III 
to V) is almost entirely devoted to the (quite long!) proof of
Theorem\,\ref{theorem1}, and is therefore mainly technical. Our
concluding comments are presented in section VI.

\section{Definitions and results}
\setcounter{equation}{0}

In order to give our results a precise and unambiguous form, we introduce the
following definitions:

\begin{enumerate}[1.]
\item A {\bf CCS-distribution} (CCS for Complete Commuting Set) in $N$
dimensions is a
  probability distribution\footnote{Here and in the following,
  probability distributions are understood as positive normalized
  measures, with an absolutely continuous part and (possibly) Dirac
  measures.} $\sigma(x_1,\ldots,x_N)$, with $x_j=q_j$ or $p_j$ for
  each index $j$.

  The CCS-distributions can occur in $2^N$ different {\bf types}, each
  {\bf type} corresponding to one choice of the $N$-tuple of arguments.
\item An {\bf $n$-chain} is a set $\{\sigma_1,\ldots,\sigma_n\}$ of
  {\sl mutually compatible} CCS-distributions of distinct types. Here,
  the mutual compatibility conditions\,\rf{I.2} read, for any pair
  $\{\sigma_\alpha (Y_\alpha),\sigma_\beta (Y_\beta)\}$, where
$Y_\alpha = \{ y_1, \ldots y_r, Y\} , \ Y_\beta = \{ y'_1, \ldots
y'_r, Y\}$ and $y'_j$  is the conjugate of $y_j$ ($y'_j=q_j$ or $p_j$
  according as $y_j=p_j$ or $q_j$),
\< \int dy_1\ldots dy_r\ \sigma_\alpha(y_1,\ldots, y_r, Y) =
   \int dy'_1\ldots dy'_r\ \sigma_\beta(y'_1,\ldots, y'_r, Y)\,, \label{II.1}
\>
  Thus, an $n$-chain is a possible  candidate for a set of $n$
  marginals, i.e. joint probability distributions obtained from some
  phase space probability distribution\footnote{We shall use sometimes
  the notation $\rho(\qp)$.} $\rho(q_1,\ldots,q_N,p_1,\ldots,p_N)$ by
  integrating over some of the arguments.

  The type of an $n$-chain is defined by the types of its elements.
\item An $n$-chain is {\bf admissible} if there exists at least one
  phase space probability distribution $\rho$ reproducing all the CCS
distributions of the
  $n$-chain, namely such that
\< \sigma_\alpha(x_1,\ldots,x_N) = \int   dx'_1\ldots dx'_N\ 
   \rho(q_1,\ldots,q_N,p_1,\ldots,p_N)\,, \qquad (\alpha=1,\ldots,n)
   \label{II.2}
\>
  where $x_i'$ is the conjugate of $x_i$. Eqs.\,\rf{II.2} imply
  \,\rf{II.1}. Using the notation $Z_\alpha = \{ x_1, \ldots x_N\},
  Z'_\alpha = \{x'_1, \ldots , x'_N\}$, and $d^NZ'_\alpha = dx'_1
  \ldots dx'_N$, \,\rf{II.2} can be rewritten as 
\[ \sigma_\alpha (Z_\alpha) = \int d^N Z'_\alpha \rho (\vec q, \vec p)
, \ (\alpha = 1, \ldots,n)
\]

\item An $n$-chain is a {\bf quantum $n$-chain} if there exists at
  least one quantum state described by the density operator
  $\hat{\rho}$ such that eqs.\,\rf{I.4} hold.
 
  Note that in that case the compatibility conditions \rf{II.1} are
  automatically satisfied.
\item Two CCS-distributions $\sigma_\alpha$ and $\sigma_\beta$ are
  {\bf contiguous} if they differ by the assignment of only one
  variable $x_i$ (to $q_i$ and $p_i$), namely
\[ \sigma_\alpha(x_1,\ldots,x_{i-1},q_i,x_{i+1},\ldots,x_N)
  \quad\mbox{and}\quad
  \sigma_\beta(x_1,\ldots,x_{i-1},p_i,x_{i+1},\ldots,x_N).
\] 
We call $i$ the {\bf index} of the pair $\{\sigma_\alpha,\sigma_\beta\}$.
\item To each $n$-chain we associate a {\bf graph} which is
  constructed as follows:
\begin{itemize}
\item to every CCS-distribution of the chain, associate a {\bf vertex}
  characterized by the collection of the variables of this distribution, 
\item connect two vertices by a {\bf link} if they correspond to
  contiguous CCS-distribu\-tions. We call {\bf index} of the link the
  index of the pair of contiguous CCS-distributions, and we say that
  the two vertices are contiguous.
\end{itemize}
  As usual, there are connected and disconnected graphs, tree graphs
  and graphs with loops. A graph $G$ completely determines the type of
  the associated $n$-chain, so that we can speak of a {\bf chain of type $G$}.
\item An $n$-chain and its associated graph are said to be {\bf proper}
  if no two links have the same index.

  Since there are at most $N$ possible indices, a proper graph has at
  most $N$ links, and thus, if it is connected, at most $(N+1)$
  vertices. Furthermore, a graph with a loop cannot be
  proper. Therefore, {\sl a connected proper graph is necessarily a tree
  graph with at most $(N+1)$ vertices.}
\item A graph $G$ is {\bf fully admissible} if all $n$-chains of type
  $G$ are admissible. 

  A graph $G$ is {\bf quantum admissible} if all quantum $n$-chains of
  type $G$ are admissible.

  Full admissibility entails quantum admissibility.

  A graph $G$ is {\bf non admissible} if it is not quantum (and a
  fortiori not fully) admissible.
\item Let a non connected graph $G$ be subgraph of a connected graph
  $G_c$. We call {\bf insertions} the vertices of $G_c$ which are not
  vertices of $G$. $G_c$ is called {\bf $G$-simple} if all its
  insertions have only two legs.
\end{enumerate}

In the following, in order to shorten the writing, particularly when drawing
graphs, we often replace the variables $q_j$ and $p_j$ by the indices
$j$ and $j'$ respectively. For example, to a CCS-distribution
$\sigma(q_1,p_2,q_3)$ we associate the vertex \rectbox{.24}{12$'$3} in a
rectangle box. As for the inserted vertices, we use instead round
boxes, e.g. \rndbox{22}{\ 12$'$3\ }.

We are now in a position to give a complete characterization of the graphs, or
equivalently of the types of $n$-chains, according to their (full, quantum
or non) admissibility.

\newtheorem{theorem}{Theorem}
\begin{theorem}{\ }
\begin{enumerate}[1.]
\item If a graph $G$ is proper and connected, then it is fully admissible.
\item If $G$ is proper but non connected, then
  \begin{enumerate}[a.]
  \item if $G$ is subgraph of a proper connected graph $G_c$, then
    \begin{enumerate}[i.]
    \item if $G_c$ is $G$-simple, $G$ is fully admissible,
    \item if $G_c$ is not $G$-simple, $G$ is quantum, but not fully,
    admissible.
    \end{enumerate}
  \item if $G$ is not subgraph of a proper connected graph $G_c$, then it
  is non admissible. 
  \end{enumerate}
\item If $G$ is non proper, then it is non admissible.
\end{enumerate}
\label{theorem1}
\end{theorem}
\kern -3mm
This theorem is complemented by the explicit construction of all the
phase space distributions $\rho$ reproducing a given chain of type
$G$, a construction which is completed only in the case of full
admissibility, that is to say when $G$ is either connected and proper,
or subgraph of a connected, proper and $G$-simple graph $G_c$ (see
section IV-B). In the case of quantum admissibility, when the
connected graph $G_c$ is proper but not $G$-simple, the situation is
not as favorable and the general expression of $\rho$ is not known
(see section V-B-2).

\kern 3mm
\noindent{\sl Remarks:}
\begin{enumerate}[1.]
\item Given a graph $G$, a proper $G_c$, when there is one (case
  2.a), is in general not unique. It is a consequence of the theorem
  that either all the proper $G_c$'s are $G$-simple, or none of them is.

  Notice that this is a pure graph theoretic statement.
\item  For $N=2$, the main result of section IV of (I), which was
  derived from Bell-like correlation inequalities, is the following :
  both in the classical and quantum cases, there exist 4-chains
  $\sigma_1(q_1,q_2),\sigma_2(p_1,q_2),\sigma_3(q_1,p_2)$ and
  $\sigma_4(p_1,p_2)$ which cannot be reproduced as marginals of any
  probability distribution $\rho(\qp)$. In the language of the present
  paper, this can be rephrased as :

{\sl The non proper graph\ \ 
\beginpicture
\setcoordinatesystem units <.1in,.1in>
\put {\rectbox{.2}{12}} [rb] at 0 1.6
\put {\rectbox{.2}{12$'$}} [lb] at 1.7 1.6
\put {\rectbox{.2}{1$'$2}} [rt] at 0 -1.6
\put {\rectbox{.2}{1$'$2$'$}} [lt] at 1.7 -1.6
\putrule from 0 2.15 to 1.7 2.15
\putrule from 0 -1.05 to 1.7 -1.05
\putrule from -1.25 1.35 to -1.25 -.2
\putrule from 3 1.35 to 3 -.2
\put {\sevenrm 2} at .85 2.65
\put {\sevenrm 2} at .85 -.55
\put {\sevenrm 1} at -0.8 .55
\put {\sevenrm 1} at 3.45 .55
\endpicture
\ \ is non admissible}.

This is the simplest case of part 3 of the above theorem, and actually it is
the clue of its proof.
\end{enumerate}

\def\figa{
\beginpicture
\setcoordinatesystem units <.1in,.1in>
\setplotarea x from -2.095 to 15.095, y from 3 to -5.9
\put {\rectbox{.35}{1234}} at 0 .345
\put {\rectbox{.35}{1$'$234}} at 6.5 .345
\put {\rectbox{.35}{1$'$2$'$34}}  at 13 .345
\put {\rectbox{.35}{12$'$3$'$4$'$}} at 0 -2.555
\put {\phantom{a}} at 0 -4
\putrule from 4.52 .9 to 1.95 .9
\putrule from 8.5 .9 to 11 .9
\put {\sevenrm 1} [b] at 3.2 1.12
\put {\sevenrm 2} [b] at 9.7 1.12
\endpicture}

\def\figb{
\beginpicture
\setcoordinatesystem units <.1in,.1in>
\setplotarea x from -8.595 to 9.2, y from 3 to -9.9
\put {\rectbox{.35}{1234}}  at -6.5 .345
\put {\rectbox{.35}{1$'$234}} at 0 .345
\put {\rectbox{.35}{1$'$2$'$34}} at 6.5 .345
\put {\rndbox{30.2}{12$'$34}} at 0 -2.71
\put {\rndbox{30.2}{12$'$3$'$4}} at 0 -5.91
\put {\rectbox{.35}{12$'$3$'$4$'$}} at 0 -9.555
\put {\phantom{a}} at 0 -11
\putrule from -4.555 .9 to -1.95 .9
\putrule from 4.55 .9 to 1.95 .9
\setlinear \plot -6.5 0.15 -2.095 -2.524 /
\setlinear \plot 6.5 0.15 2.095 -2.524 /
\putrule from 0 -3.4 to 0 -5
\putrule from 0 -6.6 to 0 -8.2
\put {\sevenrm 1} [b] at -3.2 1.12
\put {\sevenrm 2} [b] at 3.2 1.12
\put {\sevenrm 1} at 4.5 -1.8
\put {\sevenrm 2} at -4.5 -1.8
\put {\sevenrm 3} at .5 -4.15
\put {\sevenrm 4} at .5 -7.35
\endpicture}

\def\figc{
\beginpicture
\setcoordinatesystem units <.1in,.1in>
\setplotarea x from -2.095 to 15.595, y from 3 to -9.9
\put {\rectbox{.35}{1234}} at 0 .345
\put {\rectbox{.35}{1$'$234}} at 6.5 .345
\put {\rectbox{.35}{1$'$2$'$34}}  at 13 .345
\put {\rndbox{30.2}{123$'$4}} at 0 -2.71
\put {\rndbox{30.2}{123$'$4$'$}} at 0 -5.91
\put {\rectbox{.35}{12$'$3$'$4$'$}} at 0 -9.555
\put {\phantom{a}} at 0 -11
\putrule from 4.55 .9 to 1.95 .9
\putrule from 8.44 .9 to 11.05 .9
\putrule from 0 0.12 to 0 -1.85
\putrule from 0 -3.4 to 0 -5
\putrule from 0 -6.6 to 0 -8.2
\put {\sevenrm 1} [b] at 3.2 1.12
\put {\sevenrm 2} [b] at 9.7 1.12
\put {\sevenrm 3} at .5 -.85
\put {\sevenrm 4} at .5 -4.15
\put {\sevenrm 2} at .5 -7.35 
\endpicture}

\begin{figure}[!hb]
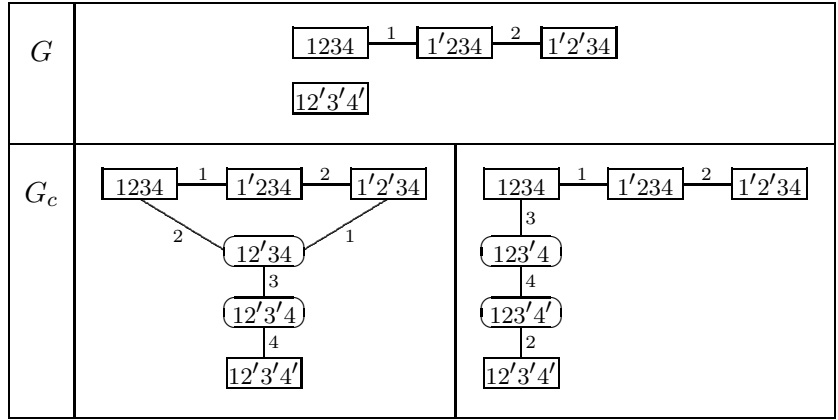

\[
\begin{tabular}{|c|c|c|}
\hline
\multicolumn{1}{|c|}{$G$} & \multicolumn{2}{c|}{\figa} \\
\hline 
{$G_c$} & {\figb} & {\figc} \\
\hline
\end{tabular}
\]
\caption{\tenrm\sl A non connected proper graph $G$ which is not subgraph of
any proper connected graph $G_c$ : $G$ is non admissible. Two
possible (non proper) $G_c$'s are shown : one with a loop, the other
one a tree graph.}
\label{figabc}
\end{figure}

\def\figd{
\beginpicture
\setcoordinatesystem units <.1in,.1in>
\setplotarea x from -2.095 to 15.595, y from 3 to -9.9
\put {\rectbox{.35}{1234}} at 0 .345
\put {\rectbox{.35}{1$'$234}} at 6.5 .345
\put {\rectbox{.35}{1$'$2$'$34}}  at 13 .345
\put {\rectbox{.35}{1$'$23$'$4$'$}} at 0 -2.955
\put {\phantom{a}} at 0 -4.5
\putrule from 4.52 .9 to 1.95 .9
\putrule from 8.455 .9 to 11.05 .9
\put {\sevenrm 1} [b] at 3.2 1.12
\put {\sevenrm 2} [b] at 9.7 1.12
\endpicture}

\def\fige{
\beginpicture
\setcoordinatesystem units <.1in,.1in>
\setplotarea x from -2.095 to 15.595, y from 3 to -9.9
\put {\rectbox{.35}{1234}} at 0 .345
\put {\rectbox{.35}{1$'$234}} at 6.5 .345
\put {\rectbox{.35}{1$'$2$'$34}}  at 13 .345
\put {\rectbox{.35}{1$'$23$'$4$'$}} at 0 -2.955
\put {\rndbox{30.2}{1$'$23$'$4}} at 6.5 -2.45
\put {\phantom{a}} at 0 -4.5
\putrule from 4.52 .9 to 1.95 .9
\putrule from 8.455 .9 to 11.05 .9
\putrule from 6.5 0.2 to 6.5 -1.56
\putrule from 4.405 -2.4 to 1.96 -2.4
\put {\sevenrm 1} [b] at 3.2 1.12
\put {\sevenrm 2} [b] at 9.7 1.12
\put {\sevenrm 3} at 6.9 -.72
\put {\sevenrm 4} [b] at 3.2 -2.177
\endpicture}

\def\figf{
\beginpicture
\setcoordinatesystem units <.1in,.1in>
\setplotarea x from -2.1 to 2.2, y from 3 to -9.9
\put {\rectbox{.22}{1$'$23}} at 0 .35
\put {\rectbox{.22}{12$'$3}} at 0 -2.95
\put {\rectbox{.22}{123$'$}} at 0 -6.25
\put {\phantom{a}} at 0 -7.8
\endpicture}

\def\figg{
\beginpicture
\setcoordinatesystem units <.1in,.1in>
\setplotarea x from -7.6 to 2.2, y from 3 to -9.9
\put {\rectbox{.22}{1$'$23}} at 0 .35
\put {\rectbox{.22}{12$'$3}} at 0 -2.95
\put {\rectbox{.22}{123$'$}} at 0 -6.25
\put {\rndbox{19}{\strut123}} at -5.5 -2.3
\put {\phantom{a}} at 0 -7.8
\setlinear \plot -4.5 -1.57 -1.315 .972 /
\setlinear \plot -4.5 -3.12 -1.315 -5.645 /
\putrule from -4.2 -2.35 to -1.31 -2.35
\put {\sevenrm 1} [b] at -3.4 -.3
\put {\sevenrm 2} [b] at -2.9 -2.15
\put {\sevenrm 3} [b] at -2.6 -4.15
\endpicture}

\begin{figure}[!ht]
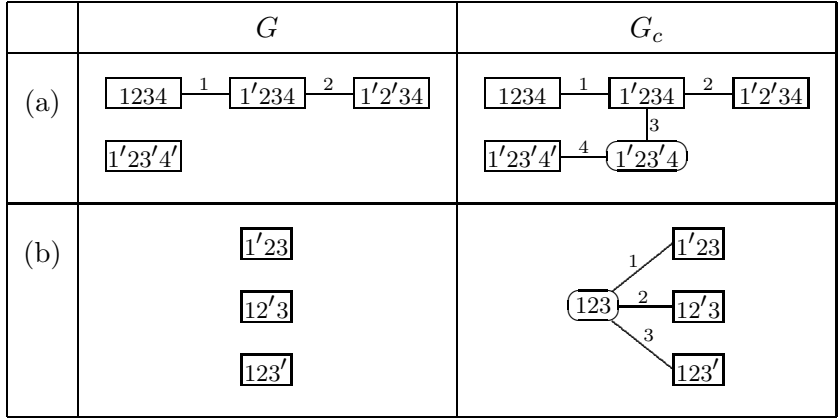

\[ \begin{tabular}{|c|c|c|}
\hline
& $G$ & $G_c$ \\
\hline
(a) & {\figd} & {\fige} \\
\hline
(b) & \figf & \figg \\
\hline
\end{tabular}
\]
\caption{(a) a non connected proper graph $G$ which is subgraph
of a proper connected $G$-simple graph $G_c$ : $G$ is fully
admissible. (b) a non connected proper graph $G$ which is subgraph
of a proper connected non $G$-simple graph $G_c$ : $G$ is quantum but
not fully admissible.}
\label{figdefg}
\end{figure}

An immediate corollary of Theorem\,\ref{theorem1} is
\begin{theorem}[$N+1$ Marginal Theorem]{\ }

A necessary condition for all quantum $n$-chains of a given type to be
admissible is $n\leq N+1$.
\label{theorem2}
\end{theorem}

It suffices to use parts 2.b and 3 of Theorem\,\ref{theorem1} and to note
that $G$ cannot have more vertices than $G_c$, and that a proper $G_c$
has at most $N+1$ vertices (see the remark after the above definition
7 of {\sl proper} $n$-chains and graphs).

Notice that, in contradistinction with the {\sl three} marginal
theorem of (I), the above theorem gives only a necessary
condition. This is because only proper connected graphs $G_c$ are
involved when $N=2$ and $n\leq N+1$, whereas non proper connected
$G_c$'s do appear as soon as $N\geq3$. Of course, for proper connected
graphs $G$ for which $n$ must be $\leq N+1$, part 1 of
Theorem\,\ref{theorem1} guarantees admissibility for $N \geq 3$ also.  

Theorem\,\ref{theorem1} will be proved in sections III to V. In order
to help understanding its content, we give in
Figs. \ref{figabc} and \ref{figdefg} examples of the various cases
encountered.

\section{Non proper graphs}
\setcounter{equation}{0}

Consider a non proper graph $G$. By definition, there exist at least
two links with the same index (say 1) connecting a first pair of
vertices $(V,V')$ and a second pair $(W,W')$. Then there necessarily
exists a second index (say 2) such that the variables $x_1$ and $x_2$
have in the four vertices $V$, $V'$, $W$ and $W'$ the assignments as
shown in Fig.\,\ref{U}.

\def\figU{
\beginpicture
\setcoordinatesystem units <.1in,.1in>
\put {\rectbox{.26}{12...}} [rb] at 16.8 0
\put {\rectbox{.26}{1$'$2...}} [rb] at 21.8 0
\put {\rectbox{.26}{12$'$...}} [rb] at 27.8 0
\put {\rectbox{.28}{1$'$2$'$...}} [rb] at 33 0
\putrule from 16.8 .523 to 18.7 .523
\putrule from 27.8 .523 to 29.7 .523
\put {$\scriptscriptstyle 1$} [b] at 28.8 -.3
\put {$\scriptscriptstyle 1$} [b] at 17.8 -.3
\put {$\scriptstyle V$} [b] at 15.4 1.7
\put {$\scriptstyle V'$} [b] at 20.3 1.7
\put {$\scriptstyle W$} [b] at 26.3 1.7
\put {$\scriptstyle W'$} [b] at 31.2 1.7
\endpicture}

\begin{figure}[thb]
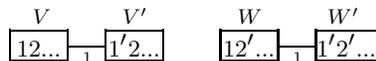

\centerline{\figU}
\caption{Critical quartet in a non proper graph.\label{U}}
\end{figure}

Quite generally, in a graph $G$ (proper or not), a set of four vertices
where a pair of variables takes the four possible assignments will be
called a {\sl critical quartet}.

We now prove
\begin{lemma}{\ }
A graph $G$ containing a critical quartet is non admissible.
\label{lemma1}
\end{lemma}
Given an $n$-chain
$\{\sigma_\alpha\}_{\alpha=1,\ldots,n}$ of type $G$ in $N$ dimensions 
and a partition of the set of indices  $\{1,\ldots,N\}= J\bigcup K$,
let us introduce the distributions
\[ \widetilde{\sigma}_\alpha(X_J)=\int\prod_{k\in K}dx_k\,
   \sigma_\alpha(X_J,X_K)\,.
\]
Some of these $\widetilde{\sigma}_\alpha$'s may coincide. We call
{\sl J-reduced $n'$-chain} the maximal set of $n'$ distinct
$\widetilde{\sigma}_\alpha$'s ($n'\leq n$). Obviously, a necessary
condition for the $n$-chain to be  admissible is that the associated
J-reduced $n'$-chain be admissible. Now, consider a quantum $n$-chain
constructed with a factorized wave function of the form
\< \Psi(q_1,\ldots,q_N)=\Psi_1(Q_J)\,\Psi_2(Q_K)\,.
\>
Choosing then $J=\{1,2\}$, to be the indices of a critical quartet, 
our results in (I) (see Remark\,2 after Theorem\,\ref{theorem1})
immediately imply the non-admissibility of the $J$-reduced 4-chain.
Hence, the non-admissibility of the $n$-chain $\sigma_\alpha$ itself,
and Lemma 1 follows.

This establishes part 3 of Theorem\,\ref{theorem1}, namely that a non proper
graph $G$ is non admissible.

\section{Connected proper graphs}
\setcounter{equation}{0}
Let $G$ be a connected proper graph. We establish part 1 of
Theorem\,\ref{theorem1} by associating to any chain of type $G$ a
particular phase space distribution $\rho_0$ reproducing this
chain. The explicit construction of such a $\rho_0$ is described in
section IV-A. In section IV-B, we derive the expression of the most
general phase space distribution reproducing the given chain.

\subsection{Particular solution}
Let $\C_n=\{\sigma_1,\ldots,\sigma_n\}$ be an $n$-chain of type
$G$. Since $G$ is a proper tree graph with $(n-1)$ links, there are
exactly $(N-n+1)$ variables which have the same assignments in all the
distributions $\sigma_\alpha$. After a possible renumbering of the
indices of the $x_i$'s, we can therefore assume that in
$\sigma_\alpha(x_1,\ldots,x_{n-1},x_n,\ldots,x_N)$ the assignment of
each of the variables $x_n$,$\ldots$,$x_N$ is independent of
$\alpha$. These variables, which will play a purely passive role, are
henceforth denoted collectively by $T$, whereas $T'$ will stand for
the set of conjugate variables $\{x'_n,\ldots,x'_N\}$.

The solution $\rho_0(q_1,\ldots,q_N,p_1,\ldots,p_N)$ of eqs.\,\rf{II.2} is
constructed as the product of ``vertex functions", ``propagators" and
an arbitrary positive function of $T'$. The former
elements are defined by the following ``Feynman rules":
\begin{enumerate}[1)]

\item to each vertex \rectfbox{$x_1,\ldots,x_N$} of $G$, we associate
the {\bf vertex function} $\sigma_\alpha(x_1,\ldots,x_N)$ of the chain $\C_n$,

\item for each link $l_i$ of $G$ carrying the index $i$, by using the
compatibility condition \rf{II.1} for the pair
$\{\sigma_{\alpha_i},\sigma_{\beta_i}\}$ of contiguous CCS-distributions
attached to this link, we define the integrated distribution
\<\begin{array}{rcl}
  \sigma_{\alpha_i\beta_i}(x_1,\ldots,x_{i-1},x_{i+1},\ldots,x_N) &=&
  \displaystyle\int dq_i\,\sigma_{\alpha_i}(x_1,\ldots,x_{i-1},q_i,
  x_{i+1},\ldots,x_N)\,,\\
  &=&\displaystyle\int dp_i\,\sigma_{\beta_i}(x_1,\ldots,x_{i-1},p_i,
  x_{i+1},\ldots,x_N)\,.
\end{array} \label{IV.1} \>
Then, to the link $l_i$ we associate the {\bf propagator}
\< \varpi_{i}(x_1,\ldots,x_{i-1},x_{i+1},\ldots,x_N) \equiv
  \left\{ \begin{array}{l}
  \displaystyle{\strut 1\over\strutup
  \sigma_{\alpha_i\beta_i}(x_1,\ldots,x_{i-1},x_{i+1},\ldots,x_N)} \\
  \quad\quad \,\mbox{  if } (x_1,\ldots,x_{i-1},x_{i+1},\ldots,x_N)\in
  \S_{\alpha_i\beta_i}\,, \\
  0 \quad\quad {\rm otherwise}\,, \end{array}\right.
  \label{IV.2}
\>
where $\S_{\alpha_i\beta_i}$ is the (essential) support of $\sigma_{\alpha_i\beta_i}$.
\end{enumerate}

The support properties of the $\sigma_\alpha$'s, $\sigma_{\alpha\beta}$'s and
$\rho_0$, and the relations between them (due to compatibility and
positivity) are not innocent in the forthcoming considerations, and we should
pay attention to them. However, doing so leads to cumbersome technicalities
which are in fact straightforward  generalizations of those developed in the
rigorous proof of Theorem 1 in (I) for the case $N=2$. Thus, in this
section IV, we shall ignore such inessential complications.

\noindent The function $\rho_0$ is then written as
\<
  \rho_0=\left(\prod_{\alpha=1}^n\sigma_\alpha\,
  \prod_{i=1}^{n-1}\varpi_i\right)\,\zeta\,,
\label{IV.03}\>
where $\zeta(T')$ is an arbitrary non negative function in
$L^1(\R^{N-n+1},\,d^{N-n+1}T')$ with normalization
\<
  \int\,d^{N-n+1}T'\,\zeta(T') = 1\,. \label{IV.04}
\>
That the expression \rf{IV.03} for $\rho_0$ solves the equations
\rf{II.2} results from the following property:

\hangafter=0 \hangindent=15pt{
Let $\widehat{V}=$ \rectfbox{$x_1,\ldots,x_N$} \,be a one-leg vertex of the
proper tree graph $G$ and $i$ be the index of the link $\hat{l}$ attached to
it. Let $\hat{\sigma}(x_1,\ldots,x_N)$ be the corresponding element of the
chain $\C_n$. Then
\<\begin{array}{rcl}
  \int dq_i\,\rho_0^{(n)}(\qp) &=&
  \rho_0^{(n-1)}(q_1,\dots,q_{i-1},q_{i+1},\ldots,q_N,p_1,\ldots,p_N) \\
  && \hfill \mbox{if  } x_i=q_i\,, \\
  \int dp_i\,\rho_0^{(n)}(\qp) &=&
  \rho_0^{(n-1)}(q_1,\ldots,q_N,p_1,\dots,p_{i-1},p_{i+1},\ldots,p_N) \\
  && \hfill\mbox{if  } x_i=p_i\,,
\end{array} \label{IV.3}
\>
where $\rho_0^{(n)}$ is the above defined $\rho_0$ associated to the $n$-chain
$\C_n$, and $\rho_0^{(n-1)}$ is the $\rho_0$ similarly associated to the
($n$-1)-chain $\C_{n-1}$ obtained from $\C_n$ by removing the CCS-distribution
$\hat{\sigma}$. Notice that the reduced chain $\C_{n-1}$ corresponds to the
reduced graph $G^{(n-1)}$ obtained from $G$ by removing the one-leg vertex
$\widehat{V}$ and the link $\hat{l}$. Hence $\rho_0^{(n-1)} =
\rho_0/(\widehat{\sigma} \varpi_i)$.}

\noindent Equation \rf{IV.3} holds because: $i$) in $\rho_0^{(n)}$ the
variable $x_i$
appears only in the factor $\hat{\sigma}$, $ii$) the propagator of the link
$\hat{l}$ is precisely the inverse of the integral of $\hat{\sigma}$ over
$x_i$.

Now, given any $\sigma_\alpha$ in the chain $\C_n$, corresponding
to the vertex $V_\alpha$ of $G$, one can start the reduction
$G^{(n)}\rightarrow G^{(n-1)}$ of the tree graph $G^{(n)}\equiv G$ at some
arbitrarily chosen one-leg vertex $\widehat{V}\neq V_\alpha$, and repeat it
$(n-1)$ times in such a way that one is left with the graph $G^{(1)}$
consisting solely of the vertex $V_\alpha$. To this ``peeling process"
$G^{(n)}\rightarrow G^{(n-1)}\rightarrow\ldots\rightarrow G^{(1)}$ of tree
graphs is naturally associated, via eqs.\,\rf{IV.3}, a reduction $\rho_0\equiv
\rho_0^{(n)}\rightarrow \rho_0^{(n-1)}\rightarrow \ldots\rightarrow
\rho_0^{(1)}$ of functions $\rho_0^{(m)}$, which eventually produces
\<
  \rho_0^{(1)}(x_1,\ldots,x_{n-1},T,T') = \int dx_1'\ldots dx_{n-1}'\,\rho_0
  = \sigma_\alpha(x_1,\ldots,x_{n-1},T)\,\zeta(T')\,, \label{IV.06}
\>
and hence, by integrating over $T'$ :
\<\int dx_1'\ldots dx_N'\,\rho_0 =
  \sigma_\alpha(x_1,\ldots,x_N)\,. \label{IV.4}
\>
Notice that, although the order of the repeated integrations over the
$x_i'$'s is imposed by the steps of the peeling process, this order becomes
obviously irrelevant in the above equation: the equations \rf{II.2}
are valid for $\rho=\rho_0$ indeed.

Finally, to illustrate our Feynman rules, let us take as an example
the graph $G^{(5)}$ of Fig.\,\ref{figh}. The distribution $\rho_0$
associated to any 5-chain of type $G^{(5)}$ is
\[ \begin{array}{rcl}
  \rho_0(\qp) &=& \displaystyle
  \sigma_1(q_1q_2q_3q_4)\,{1\over\sigma_{12}(q_2q_3q_4)}\,
  \sigma_2(p_1q_2q_3q_4)\,{1\over\sigma_{23}(p_1q_3q_4)}\,
  \sigma_3(p_1p_2q_3q_4) \\
  && \displaystyle
  {1\over\sigma_{24}(p_1q_2q_4)}\,\sigma_4(p_1q_2p_3q_4)\,
  {1\over\sigma_{45}(p_1q_2p_3)}\,\sigma_5(p_1q_2p_3p_4)\,.
  \end{array}
\]
Here and in the sequel, we keep writing the propagators as
$1/\sigma_{\alpha\beta}$, although they are strictly given by eq.\,\rf{IV.2}.

\def\figh{
\beginpicture
\setcoordinatesystem units <.1in,.1in>
\setplotarea x from -2.095 to 15.595, y from 3 to -9.9
\put {\rectbox{.35}{1234}} at 0 .345
\put {\rectbox{.35}{1$'$234}} at 6.5 .345
\put {\rectbox{.35}{1$'$2$'$34}}  at 13 .345
\put {\rectbox{.35}{1$'$23$'$4$'$}} at 0 -2.955
\put {\rectbox{.35}{1$'$23$'$4}} at 6.5 -2.955
\put {\phantom{a}} at 0 -4.5
\putrule from 4.505 .777 to 1.995 .777
\putrule from 8.495 .777 to 11.005 .777
\putrule from 6.5 0.1 to 6.5 -1.645
\putrule from 4.505 -2.524 to 1.995 -2.524
\put {\sevenrm 1} [b] at 3.2 1
\put {\sevenrm 2} [b] at 9.7 1
\put {\sevenrm 3} at 7 -.75
\put {\sevenrm 4} [b] at 3.2 -2.3
\endpicture}

\begin{figure}[ht]
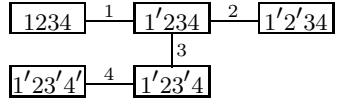

\centerline\figh
\caption{a proper connected graph with $N$=4 and $n$=5.}
\label{figh}
\end{figure}

\subsection{General solution}
Let us define the $n$ following positive measures, each one associated to a
particular component $\sigma_\alpha$ of the $n$-chain $\C_n$
\< d\mu_\alpha =
  \left\{ \begin{array}{ll}
  \displaystyle{\strutdown\rho_0(\vec q, \vec p)\over
  \strutup\sigma_\alpha(Z_\alpha)}\ d^N Z'_\alpha \qquad
  & \mbox{if  } Z_\alpha = (x_1,\ldots,x_N)\in\S_\alpha\,, \\
  0 & \mbox{otherwise}\,,
  \end{array} \right. \label{IV.6}
\>
where $\S_\alpha$ denotes the (essential) support of $\sigma_\alpha$. Due
to eq.\,\rf{IV.4} these $(x_1,\ldots,x_N)$-dependent
measures\footnote{Notice also that the $d\mu_\alpha$'s are continuous
linear mappings of 
$L^1(\R^{2N},\rho_0\,d^{\kern .05em N}\!q\,d^{\kern .05em N}\!p)$
into itself.} are normalized for all $(x_1,\ldots,x_N)\in\S_\alpha$.

It is convenient to write the general solution $\rho$ we are looking for in
the form
\<\rho=\rho_0\,(1+\lambda\,h)\,. \label{IV.7}
\>
Here, the function $h(\qp)$ will be chosen as to ensure
eq.\,\rf{II.2}, which results in the linear equations 
\<\int d^N Z'_\alpha \,\rho_0(\qp)\,h(\qp)=0 \, ,\qquad\qquad
  (\alpha=1,\ldots,n)\,, \label{IV.8}
\>
whereas the real constant $\lambda$ is a normalisation factor which will be
useful to control the positivity of $\rho$. Thanks to definitions \rf{IV.6},
eqs.\,\rf{IV.8} are equivalent to
\< \int d\mu_\alpha\,h=0 \qquad\qquad (\alpha=1,\ldots,n)\,. \label{IV.9}
\>
We then observe that, for any $\alpha$ and any function $g$ in
$L^1(\R^{2N},\rho_0\,d^{\kern .05em N}\!q\,d^{\kern .05em N}\!p)$ 
\< \int d\mu_\alpha\,\left(\int d\mu_\alpha\,g\right)=\int d\mu_\alpha\,g\,,
  \label{IV.10}
\>
since $\int d\mu_\alpha\,g$ does not depend any longer on the
integration variables $(x_1',\ldots,x_N')$ and $d\mu_\alpha$ is
normalized. That is, the linear operators $P_\alpha$ : 
$L^1(\R^{2N},\rho_0\,d^{\kern .05em N}\!q\,d^{\kern .05em N}\!p)
\rightarrow
L^1(\R^{2N},\rho_0\,d^{\kern .05em N}\!q\,d^{\kern .05em N}\!p)$
defined by $P_\alpha\,g = \int d\mu_\alpha\,g$ are projectors,
\<
  P_\alpha^2 = P_\alpha \qquad\qquad\qquad (\alpha=1,\ldots,n)\,.
  \label{IV.013}
\>
The set $\{P_\alpha\}$ enjoys certain algebraic properties which are
crucial for the construction of the general solution $h$ of
eqs.\,\rf{IV.9} :
\begin{lemma}{\ }
  \begin{enumerate}[a)]
  \item The projectors $P_\alpha$ and $P_\beta$ associated with any pair
    $\{V_\alpha,V_\beta\}$ of contiguous vertices commute
  \<
    [P_\alpha,P_\beta] = 0\,. \label{IV.014}
  \>
  \item If $V_\alpha$, $V_\beta$ and $V_\gamma$, are three   vertices
  of  the connected proper (tree) graph $G$ such that $V_\alpha$
  belongs  to the (unique) path connecting $V_\beta$ to $V_\gamma$,
  and is contiguous to at least one of these two vertices, then
  \<
    P_\gamma\,P_\alpha\,P_\beta = P_\gamma\,P_\beta\,. \label{IV.015}
  \>
  \end{enumerate}
\label{lemma2}
\end{lemma}
The proof is given in \ref{Appendix lemma2} We stress that the contiguity of
$V_\alpha$ with $V_\beta$ or $V_\gamma$ is essential for the validity of
eq.~\rf{IV.015}.

Let us now introduce the central object of our construction, namely
the operator
\< \Pi = 1-\sum_{\alpha=1}^n P_\alpha +
   \sum_{i=1}^{n-1}\,P_{\alpha_i}\,P_{\beta_i}\,, \label{IV.15}
\>
where $P_{\alpha_i}$ and $P_{\beta_i}$ denote the operators $P$
associated with the two (contiguous) vertices $V_{\alpha_i}$ and
$V_{\beta_i}$ attached to the link with index $i$. Thanks to
Lemma~\ref{lemma2}, it is readily shown that $\Pi$ is annihilated by
all the projectors $P_\alpha$ :
\<
  P_\gamma\,\Pi = 0 \qquad\qquad\qquad (\gamma=1,\ldots,n)\,. \label{IV.017}
\>
Indeed :
\<\begin{array}{rcl}
  P_\gamma\,\Pi &=& P_\gamma-\sumd_{\alpha=1}^n P_\gamma\,P_\alpha
  +\sumd_{i=1}^{n-1} P_\gamma\,P_{\alpha_i}\,P_{\beta_i}\,, \\
  &=& -P_\gamma\,
  \sumd_{\buildrel{\scriptstyle\alpha=1}\over{\alpha\neq\gamma}}^n\,
  P_\alpha + \sumd_{i=1}^{n-1}\,P_\gamma\,P_{\alpha_i}\,P_{\beta_i}\,.
  \end{array} \label{IV.018}
\>
But, according to eqs.~\rf{IV.015} :
\<
  P_\gamma\,P_{\alpha_i}\,P_{\beta_i} = P_\gamma\,P_{\delta_i}\,,
\>
where $\delta_i=\beta_i$ (resp. $\alpha_i$) if $V_{\alpha_i}$ 
(resp. $V_{\beta_i}$) belongs to the path connecting $V_\gamma$ to
$V_{\beta_i}$ (resp. $V_{\alpha_i}$). Hence
\<
  \sum_{i=1}^{n-1}\,P_\gamma\,P_{\alpha_i}\,P_{\beta_i} = P_\gamma\,
  \sum_{\buildrel{\scriptstyle\alpha=1}\over{\alpha\neq\gamma}}^n\,
  P_\alpha\,,\label{IV.020}
\>
which entails eq.~\rf{IV.017} (the property $\Pi\,P_\gamma=0$
($\gamma=1,\ldots,n$), which also holds as a consequence of
eq.~\rf{IV.015}, will not be used here). Note that eq.\,\rf{IV.020}
would not be valid if the graph $G$ were not connected.

Furthermore, $\Pi$ is itself a projector :
\<
  \Pi^2 = \Pi\,, \label{IV.021}
\>
as immediately deduced from
\[
  \Pi^2 = (1-\sum_{\alpha=1}^n\,P_\alpha +
  \sum_{i=1}^{n-1}\,P_{\alpha_i}\,P_{\beta_i})\,\Pi
\]
and eq.~\rf{IV.017}. This operator allows us to write down at once the
general solution of eqs.~\rf{IV.9}, i.e.
\<
  P_\alpha\,h = 0 \kern 3cm (\alpha=1,\ldots,n)\,,\label{IV.022}
\>
as
\<
  h = \Pi\,f\,, \label{IV.023}
\>
where $f$ is an arbitrary function in $L^1(\R^{2N},d^{\kern .05em
  N}\!q\,d^{\kern .05em N}\!p)$. That eq.~\rf{IV.023} implies
eqs.~\rf{IV.022} is trivial due to eq.~\rf{IV.017}. Conversely, any
function $h$ satisfying eqs.~\rf{IV.022} assumes the form \rf{IV.023} :
since then $h=\Pi\,h$, it suffices to take $f=h$.

We now have to give the representation formula for $h$ resulting from
eqs.\,\rf{IV.023} and \rf{IV.15} an explicit form in terms of the
data of the problem, namely the elements of the chain $\C_n$. For this
purpose, it is necessary to use appropriate notations. First, we
denote by $Z_\alpha$ the collection of arguments of the vertex
function $\sigma_\alpha$, and $Z_\alpha'$ the collection of the
conjugate arguments (a notation already used in the definition
\rf{IV.6}). Then
\< (P_\alpha f)(Z_\alpha) = {1\over\sigma_\alpha(Z_\alpha)} 
  \int d^{\kern .05em N}\!Z_\alpha'\,\rho_0(\qp)\,f(\qp)\,. \label{IV.180}
\>
Second, we denote by $\sigma_{\alpha_i}(X_i,x_i)$ and
$\sigma_{\beta_i}(X_i,x_i')$ the vertex functions of the vertices
$V_{\alpha_i}$ and $V_{\beta_i}$, where
$X_i=\{x_1,\ldots,x_{i-1},x_{i+1},\ldots,x_N\}$. Accordingly, we write
$\rho_0(X_i,X_i',x_i,x_i')$ for $\rho_0(\qp)$, where
$X_i'=\{x_1',\ldots,x_{i-1}',x_{i+1}',\ldots,x_N'\}$, and so on. With
these notations
\<(P_{\alpha_i}P_{\beta_i}\,f)(X_i) = \int d^{\kern .05em N-1}\!X_i'\,dx_i'\,
  {\rho_0(X_i,X_i',x_i,x_i')\over\sigma_{\alpha_i}(X_i,x_i)}\,
  (P_{\beta_i}\,f)(X_i,x_i')\,, \label{IV.18}
\>
and
\<(P_{\beta_i}\,f)(X_i,x_i') = \int d^{\kern .05em N-1}\!X_i'\,dx_i\,
  {\rho_0(X_i,X_i',x_i,x_i')\over\sigma_{\beta_i}(X_i,x_i')}\,
  f(X_i,X_i',x_i,x_i')\,. \label{IV.19}
\>
In the r.h.s. of eq.\,\rf{IV.18}, one observes that the integral $\,\int\!
d^{\kern .05em N-1}\!X_i'\,\rho_0\,$ can be performed explicitly by
means of the ``peeling process" described in section IV-A\,:
\<\int d^{\kern .05em N-1}\!X_i'\ \rho_0(X_i,X_i',x_i,x_i') =
  {\sigma_{\alpha_i}(X_i,x_i)\,\sigma_{\beta_i}(X_i,x_i')\over
  \sigma_{\alpha_i\beta_i}(X_i)}\,. \label{IV.20}
\>
This equation obtains by stopping the peeling process at the reduced graph
made of the two vertices $V_{\alpha_i}$, $V_{\beta_i}$ and the link between
them. Here appears the propagator $1/\sigma_{\alpha_i\beta_i}$ with
\< \sigma_{\alpha_i\beta_i}(X_i) = \int dx_i\,\sigma_{\alpha_i}(X_i,x_i)
   = \int dx_i'\,\sigma_{\beta_i}(X_i,x_i')\,. \label{IV.21}
\>
Then, inserting eqs.\,\rf{IV.19} and \rf{IV.20} in eq.\,\rf{IV.18}, one gets,
after simplifications,
\< (P_{\alpha_i}P_{\beta_i}\,f)(X_i) =
  \int d^{\kern .05em N-1}\!X_i'\,dx_i\,dx_i'
  \ {\rho_0(X_i,X_i',x_i,x_i')\over\sigma_{\alpha_i\beta_i}(X_i)}\,
  f(X_i,X_i',x_i,x_i')\,. \label{IV.22}
\>
Finally, collecting eqs.\,\rf{IV.023}, \rf{IV.15}, \rf{IV.180} and
\rf{IV.22}, we obtain the expression of the function $h$ we were looking for
\< \begin{array}{rcl}
  h(\qp) &=&
  f(\qp)-\displaystyle{\sum_{\alpha=1}^n {1\over\sigma_\alpha(Z_\alpha)}}
  \int d^{\kern .05em N}\!Z_\alpha'\,\rho_0(\qp)\,f(\qp) \\
  && +\displaystyle{\sum_{i=1}^{n-1} {1\over\sigma_{\alpha_i\beta_i}(X_i)}}\,
  \int d^{\kern .05em N-1}\!X_i'\,dx_i\,dx_i'\ \rho_0(\qp)\,f(\qp)\,.
  \end{array} \label{IV.23}
\>

Equations \rf{IV.7} and \rf{IV.23} provide us with the general solution
$\rho$ of the linear system \rf{II.2}. It remains to enforce the positivity
of this solution. Let us denote by $m_+$ (resp. $-m_-$) the (essential)
supremum (resp. infimum) of $h$. Because of eq.\,\rf{IV.8}, $m_+$ and $m_-$
are strictly positive when $h$ does not vanish identically. Then, from
eq.\,\rf{IV.7}, the condition $\rho\geq0$ is equivalent to the condition on the
parameter $\lambda$
\< -{1\over m_+}\leq\lambda\leq {1\over m_-}\,. \label{IV.24}
\>
We stress that the allowed interval $[-{1\over m_+},{1\over m_-}]$ is not
zero as soon as the range of the arbitrary function $f$
is (essentially) bounded. Indeed, assuming that $A\leq f\leq B$
(almost) everywhere, one finds from eq.\,\rf{IV.23} that $m_\pm\leq n\,(B-A)$.

Equations \rf{IV.7}, \rf{IV.23} and \rf{IV.24} for $\rho$ constitute the
generalization of the results in section V of (I) (see eqs.\,(V.8) to
(V.10) there) to phase spaces of arbitrary dimension, in one case of full
admissibility (connected proper graphs).

\section{Non connected proper graphs}
\setcounter{equation}{0}

Throughout this section, devoted to the proof of part 2 of
Theorem\,\ref{theorem1}, a vertex (or insertion) with only two legs
will be called {\sl simple vertex}.

\subsection{Non proper $G_c$}

Our purpose here is to establish part 2.b of Theorem\,\ref{theorem1}.
To this end we first construct a particular connected graph $G_c$ such
that the proper graph $G$ is subgraph of $G_c$. By hypothesis $G_c$
is not proper. We then show that this implies the existence of at
least one critical quartet {\sl in the graph} $G$. According to
Lemma \ref{lemma1}, the statement 2.b immediately follows.

Let $G=\bigcup_k G_k$ be the decomposition of a proper graph $G$ into
connected components $G_k$'s. Each $G_k$ is a proper graph, hence a tree.
We connectify $G$ recursively according to the following scheme.
Assume we have already connectified the components $G_1,\ldots,G_k$
into a connected {\sl diagram} $\Gamma_k$. We define $\Gamma_{k+1}$ as
follows. Let us call {\sl segment} a linear chain of inserted simple
vertices and links, and define its length by the number of its
links. We choose one of the shortest segments which connect 
$\Gamma_k$ to the component $G_r$ ($r=k+1,\ldots$). Among the $G_r$'s
we select one, say $G_{k+1}$, which minimizes the length of the
attached segment. We call $\Sigma_k$ this segment. The diagram
$\Gamma_{k+1}$ is defined by
$\Gamma_{k+1}=\Gamma_k\bigcup G_{k+1}\bigcup\Sigma_k$.

\def\arrow
{\beginpicture
   \setlinear \plot -0.1 0  -1 .3  -.8 0  -1 -.3 -0.1 0 /
   \setlinear \plot 0.1 0  1 .3  .8 0  1 -.3 0.1 0 /
   \setdots <2pt>
   \putrule from 0 .1 to 0 1.6
\endpicture}

\def\rdashes
{\beginpicture
   \setlinear \plot 1.9 0  1 .3  1.2 0  1 -.3  1.9 0 /
   \setdashes <1.1pt>
   \putrule from 0 0 to 1.1 0
\endpicture}

\def\ldashes
{\beginpicture
   \setlinear \plot -.8 0  .1 .3  -.1 0  .1 -.3  -.8 0 /
   \setdashes <1.1pt>
   \putrule from 0 0 to 1.1 0
\endpicture}

\def\figX{
\beginpicture
\setcoordinatesystem units <.1in,.1in>
\put {\rndbox{20}{}} [lt] at 2.6 -.15
\put {\rectbox{.2}{}} [lt] at 7.9 -1.25
\put {\arrow} at 4 -3.145
\put {\arrow} at 7.8 -3.145
\putrule from 4.8 -4 to 7 -4
\putrule from 0 -4 to 3.2 -4
\putrule from 8.6 -4 to 12 -4
\put {\ldashes} at -.65 -4
\put {\rdashes} at 12.65 -4
\setdashes <1.52pt>
\setlinear \plot 0.1 1.1  1.2 0 /
\setlinear \plot 0.1 -2.5  1.2 -1.4 /
\setlinear \plot 4.2 -0.7  7.9 -0.7 /
\setlinear \plot 10.8 -0.7  12.7 -0.7 /
\put {$\scriptstyle\Gamma_k$} at 1.5 -5
\put {$\scriptstyle\Sigma_k$} at 6 -5
\put {$\scriptstyle G_{k+1}$} at 11.1 -5
\endpicture
}

\noindent Note that
\begin{enumerate}[1)]
\item in the above construction, two contiguous vertices of a $\Gamma$
are not necessarily linked, so that the {\sl diagrams} $\Gamma$ are
not always {\sl graphs} as defined in section II. The advantage of
this construction is that the $\Gamma_k$'s are trees.
\item the segment $\Sigma_k$ is attached to $G_{k+1}$ through a
vertex of $G$ and attached to $\Gamma_k$ through either a vertex of $G$
or an inserted vertex of $\Gamma_k$ (as represented in Fig.\,\ref{X}). In the
latter case, this inserted vertex becomes non simple (if it was simple
before).

\vspace{5mm}
\begin{figure}[thb]
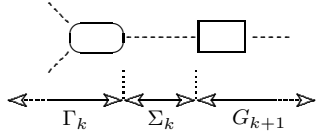

\centerline{\figX}
\caption{the diagram $\Gamma_{k+1}$ \label{X}}
\end{figure}

\item there is in general some arbitrariness in the construction of
the $\Gamma_k$'s. First, the recursive process has to be initialized
by the choice of one component as $\Gamma_1$. Next, in the
subsequent steps of the process, there is a possible arbitrariness
in the choice of $G_{k+1}$ and its attached segment.
\end{enumerate}

Once this connectification process is completed, we end up with a
connected tree diagram $\Gamma_c$ which contains all the components of
$G$. If $\Gamma_c$ is not a graph, we obtain a graph $G_c$ by adding
links between all pairs of contiguous vertices which are still not
linked in $\Gamma_c$.

$G_c$ and $\Gamma_c$ may coincide or not. In the latter case, it is important
to notice that $G_c$ and $\Gamma_c$ are still either both proper or both non
proper. This follows from the fact that i) when going from $\Gamma_c$ to
$G_c$, a loop of $G_c$ is created each time one adds a link, ii) a loop
contains at least two pairs of links carrying the same index.

Let us define the diagrams $\Omega_k$ by
\[ \Omega_k=\Gamma_k\bigcup G_{k+1}\bigcup G_{k+2}\bigcup\ldots
\]
which satisfy the inclusion relations
\[ \Omega_1\equiv G\subset\Omega_2\subset\ldots\subset
   \Omega_c\equiv\Gamma_c\subset G_c\,.
\]

Now, by hypothesis $G_c$, and thus also $\Omega_c=\Gamma_c$, are non
proper, whereas $\Omega_1=G$ is proper. This implies the existence of an
integer $k$ such that $\Omega_k\subset\Omega_{k+1}$ with $\Omega_k$
proper and $\Omega_{k+1}$ non proper. From the observation that
\[ \Omega_{k+1}=\Omega_k\bigcup\Sigma_k\,,
\]
we deduce that there exists at least one index, say $1$, which is
carried by just two links, one $l_1$ in $\Sigma_k$ and a second one
$l'_1$ in $\Omega_k$. The link $l'_1$ may appear either in $\Gamma_k$
or in the components $G_{k+1},\ldots$ of $G$, which leads us to
distinguish three cases :
\begin{enumerate}[a)]
\item $l'_1\subset G_{k+1}$,
\item $l'_1\subset\Gamma_k$,
\item $l'_1\subset G_{k+2}\mbox{ or }G_{k+3}\mbox{ or }\ldots$
\end{enumerate}

We now proceed with a few remarks which will be useful in the forthcoming
argument, although not always explicitly refered to thereby :
\begin{enumerate}[1)]
\item All the end points (one leg vertices) of the $\Gamma_k$'s belong
to $G$.
\item Any link of a $\Gamma_k$ belongs to at least one linear chain
with end vertices belonging to $G$.
\item On a segment, the indices of the links can be reordered at our
convenience. This should be kept in mind when constructing the
$\Gamma_k$'s.
\item Two links carrying the same index cannot be attached to a common
vertex.

As a consequence of these last two remarks, since all the $\Sigma_k$'s
are shortest connecting chains,
\item all segments $\Sigma_k$ are proper, and
\item on a connected tree $\Gamma_k$, the (unique) path joining two links
carrying the same index contains either two vertices of $G$, or one
vertex of $G$ and one inserted non simple vertex, or two inserted non
simple vertices.
\end{enumerate}

\noindent{\bf Case a)}

Let $V$ be any vertex of $\Gamma_k$ belonging to $G$, and $V'$ be the
vertex of $G_{k+1}$ where the segment $\Sigma_k$ attaches. Figure
\ref{Y} exhibits the relevant part of $\Omega_{k+1}$, namely the
linear chain joining the vertices $V$ and $V'$, and the linear chain
from $V'$ to the link $l'_1$ in $G_{k+1}$. Moreover, in accordance
with Remark\,3), we have attached the link $l_1$ to $V'$. We have also
called 2 the index of the other link attached to $V'$.

\def\figY{
\beginpicture
\setcoordinatesystem units <.1in,.1in>
\put {\rectbox{.2}{12...}} [rb] at 0 -0.13
\put {\rndbox{20}{}} [lt] at 4.62 1
\put {\rndbox{20}{}} [lt] at 10.62 1
\put {\rectbox{.24}{1$'$2...}} [rb] at 16.8 -0.13
\put {\rectbox{.22}{}} [rb] at 21.2 -0.13
\put {\rectbox{.28}{1$'$2$'$...}} [rb] at 27.8 -0.13
\put {\rectbox{.24}{12$'$...}} [rb] at 32.6 -0.13
\putrule from 12 .4 to 13.9 .4
\putrule from 16.8 .4 to 18.5 .4
\putrule from 27.8 .4 to 29.7 .4
\put {\arrow} at 6 -2.015
\put {\arrow} at 13.7 -2.015
\putrule from -3 -2.87 to 5.2 -2.87
\putrule from 6.8 -2.87 to 12.9 -2.87
\putrule from 14.5 -2.87 to 33 -2.87
\put {\ldashes} at -3.65 -2.87
\put {\rdashes} at 33.65 -2.87
\setdashes <1.52pt>
\putrule from 0 .4 to 3.3 .4
\putrule from 6 .4 to 9.2 .4
\putrule from 21.3 .4 to 24.5 .4
\put {$\scriptstyle l_1$} [b] at 12.9 .9
\put {$\scriptstyle l_1'$} [b] at 28.7 .6
\put {$\scriptscriptstyle 1$} [b] at 12.8 -.4
\put {$\scriptscriptstyle 1$} [b] at 28.7 -.4
\put {$\scriptscriptstyle 2$} [b] at 17.6 -.4
\put {$\scriptstyle V$} [b] at -1.3 1.7
\put {$\scriptstyle V'$} [b] at 15.5 1.7
\put {$\scriptstyle W'$} [b] at 26.3 1.7
\put {$\scriptstyle W$} [b] at 31.1 1.7
\put {$\scriptstyle\Gamma_k$} at 1.1 -3.8
\put {$\scriptstyle\Sigma_k$} at 9.8 -3.8
\put {$\scriptstyle G_{k+1}$} at 24.5 -3.8
\endpicture
}

\begin{figure}[thb]
\centerline{\figY}
\caption{Case a) \label{Y}}
\end{figure}

According to Remarks\,3) and 4), there is no link with index 2 in
$\Sigma_k$. Furthermore, since $\Gamma_k\bigcup G_{k+1}$ is proper, no
new link of index 1 or 2 can appear in the linear chain
connecting $V$ to $W$. As a consequence, the vertices $V$, $V'$, $W$
and $W'$ constitute a critical quartet.

\noindent{\bf Case b)}

The relevant part of $\Omega_{k+1}$ is displayed in Fig.\,\ref{Z}. Here, 
$V$ is a vertex of $\Gamma_k$ belonging to $G$, such that the (unique)
path joining it to $\Sigma_k$ contains the link $l'_1$.

\def\figZ{
\beginpicture
\setcoordinatesystem units <.1in,.1in>
\put {\rectbox{.2}{12...}} [rb] at 0 -0.13
\put {\rndbox{20}{}} [rt] at 4.6 1
\put {\rndbox{20}{}} [rt] at 9.2 1
\put {\rndbox{20}{}} [rt] at 15.2 1
\put {\rndbox{20}{}} [rt] at 21.2 1
\put {\rndbox{20}{}} [rt] at 25.8 1
\put {\rndbox{20}{}} [rt] at 30.4 1
\put {\rectbox{.24}{12$'$...}} [rb] at 38 -0.13
\put {\rectbox{.24}{1$'$2...}} [rb] at 16.77 3.35
\put {\rectbox{.28}{1$'$2$'$...}} [rb] at 27.48 3.35
\putrule from 6 .4 to 7.8 .4
\putrule from 22.6 .4 to 24.4 .4
\putrule from 27.2 .4 to 29 .4
\put {\arrow} at 27.3 -2.015
\put {\arrow} at 35 -2.015
\putrule from -3 -2.87 to 26.5 -2.87
\putrule from 28.1 -2.87 to 34.2 -2.87
\putrule from 35.8 -2.87 to 39 -2.87
\put {\ldashes} at -3.65 -2.87
\put {\rdashes} at 39.65 -2.87
\setdashes <1.52pt>
\putrule from 0 .4 to 3.3 .4
\putrule from 10.6 .4 to 13.9 .4
\putrule from 16.6 .4 to 19.9 .4
\putrule from 31.8 .4 to 35.1 .4
\putrule from 15.2 3.1 to 15.2 1
\putrule from 25.8 3.1 to 25.8 1
\put {$\scriptstyle l_1'$} [b] at 6.9 .6
\put {$\scriptstyle l_1$} [b] at 28.2 .9
\put {$\scriptscriptstyle 1$} [b] at 6.9 -.4
\put {$\scriptscriptstyle 1$} [b] at 28.1 -.4
\put {$\scriptscriptstyle 2$} [b] at 23.5 -.4
\put {$\scriptstyle V$} [b] at -1.3 1.7
\put {$\scriptstyle V'$} [b] at 15.2 5.2
\put {$\scriptstyle \widetilde{V}'$} [b] at 15.2 -1.7
\put {$\scriptstyle W'$} [b] at 25.8 5.2
\put {$\scriptstyle \widetilde{W}'$} [b] at 25.8 -1.7
\put {$\scriptstyle W$} [b] at 36.8 1.7
\put {$\scriptstyle\Gamma_k$} at 12 -3.8
\put {$\scriptstyle\Sigma_k$} at 31 -3.8
\put {$\scriptstyle G_{k+1}$} at 38.2 -3.8
\endpicture
}

\begin{figure}[thb]
\centerline{\figZ}
\caption{Case b) \label{Z}}
\end{figure}

The existence of the non simple vertices $\widetilde{V}'$ and
$\widetilde{W}'$ results from Remark\,6). They may possibly coincide
with respectively the vertices $V'$ and $W'$ of $G$. As previously, no
new index 1 or 2 can appear in the chain displayed in Fig.\,\ref{Z}, which
implies that the vertices $V$, $V'$, $W$ and $W'$ constitute a
critical quartet.

\noindent{\bf Case c)}

In that case, the relevant part of $\Omega_{k+1}$ is made of two
disconnected parts, as displayed in Fig.\,\ref{T}. As previously, the non
simple inserted vertex $\widetilde{V}$ may possibly coincide with $V$.

\def\figT{
\beginpicture
\setcoordinatesystem units <.1in,.1in>
\put {\rectbox{.2}{1...}} [rb] at 0 -0.13
\put {\rectbox{.2}{1...}} [lb] at 3.25 3.35
\put {\rndbox{20}{}} [lt] at 4.62 1
\put {\rndbox{20}{}} [lt] at 10.62 1
\put {\rectbox{.2}{1$'$...}} [rb] at 16.2 -0.13
\put {\rectbox{.2}{1...}} [rb] at 27.8 -0.13
\put {\rectbox{.2}{1$'$...}} [rb] at 32 -0.13
\putrule from 12 .4 to 13.7 .4
\putrule from 27.8 .4 to 29.5 .4
\put {\arrow} at 6 -2.015
\put {\arrow} at 13.7 -2.015
\putrule from -3 -2.87 to 5.2 -2.87
\putrule from 6.8 -2.87 to 12.9 -2.87
\putrule from 14.5 -2.87 to 17 -2.87
\putrule from 24.7 -2.87 to 32.7 -2.87
\put {\rdashes} at 17.65 -2.87
\put {\ldashes} at -3.65 -2.87
\put {\ldashes} at 24.05 -2.87
\put {\rdashes} at 33.35 -2.87
\setdashes <1.52pt>
\putrule from 0 .4 to 3.3 .4
\putrule from 6 .4 to 9.2 .4
\putrule from 4.6 3.1 to 4.6 1
\put {$\scriptstyle l_1$} [b] at 12.9 .9
\put {$\scriptstyle l_1'$} [b] at 28.7 .6
\put {$\scriptscriptstyle 1$} [b] at 12.8 -.4
\put {$\scriptscriptstyle 1$} [b] at 28.7 -.4
\put {$\scriptstyle V$} [b] at -1.3 1.7
\put {$\scriptstyle \widehat{V}$} [b] at 4.6 5.2
\put {$\scriptstyle \widetilde{V}$} [b] at 4.6 -1.7
\put {$\scriptstyle V'$} [b] at 15.25 1.7
\put {$\scriptstyle W$} [b] at 26.4 1.7
\put {$\scriptstyle W'$} [b] at 31 1.7
\put {$\scriptstyle\Gamma_k$} at 1.1 -3.8
\put {$\scriptstyle\Sigma_k$} at 9.8 -3.8
\put {$\scriptstyle G_{k+1}$} at 16.5 -3.8
\put {$\scriptstyle G_{k+2}$} at 28.7 -3.8
\endpicture
}

\begin{figure}[thb]
\centerline{\figT}
\caption{Case c) \label{T}}
\end{figure}

Let us denote by $I$ the set of indices appearing (each only once) in
the links between $V$ and $\widetilde{V}$, by $J$ the set of indices
appearing in the segment $\Sigma_k$ but the index 1, and by $K$ all
the remaining indices. We can assume $I\bigcap J=\emptyset$ (otherwise
the configuration would also enter the case b)). We further split the
sets $I$, $J$ and $K$ as $I=I_1\bigcup I_2$, $J=J_1\bigcup J_2$,
$K=K_1\bigcup K_2$. Here, $I_1$ and $I_2$ are introduced to separate
the variables which have the same or different assignments in the
vertices $V$ on the one hand and in the vertices $W$ and $W'$ on the
other hand, and similarly for the splitting of $J$ and $K$. Then
$\{1,I_1,I_2,J_1,J_2,K_1,K_2\}$ is a partition of $\{1,2,\ldots,N\}$,
and the assignments of the variables in the vertices $V$, $V'$, $W$
and $W'$ of $G$, and $\widetilde{V}$ take the form
\< \left\{ \begin{array}{rcl}
   V &= &\{1\,\I_1\,\I_2\,\J_1\,\J_2\,\K_1\,\K_2\} \\
   \widetilde{V} &= &\{1\,\I'_1\,\I'_2\,\J_1\,\J_2\,\K_1\,\K_2\} \\
   V' &= &\{1'\,\I'_1\,\I'_2\,\J'_1\,\J'_2\,\K_1\,\K_2\} \\
   W &= &\{1\,\I_1\,\I'_2\,\J_1\,\J'_2\,\K_1\,\K'_2\} \\
   W' &= &\{1'\,\I_1\,\I'_2\,\J_1\,\J'_2\,\K_1\,\K'_2\}
   \end{array} \right. \label{5.1}
\>
In these formulas, $\I_1,\I_2,\ldots$ represent collections of variables $q$
and $p$. In accordance with our convention (see sect.II), $\I_1$ is
written as a set of indices, namely those of $I_1$ but each one being
primed or not. As for $\I'_1$, it is written as a set of the same
indices, non primed (resp. primed) if primed (resp. non primed) in
$\I_1$. Similarly for the other sets $\I_2,\I'_2,\ldots$

Let us define the distance $d(U,U')$ between two vertices $U$ and $U'$
as the number of variables with different assignments in $U$ and
$U'$. Inspecting eq.\,\rf{5.1}, one readily obtains
\< \left\{ \begin{array}{rcl}
   d(\widetilde{V},V') &= &1+j_1+j_2\,, \\
   d(\widetilde{V},W) &= &i_1+j_2+k_2\,,
   \end{array} \right. \label{5.2}
\>
where $j_1=\mbox{card}\ J_1$, and so on. Since $\Sigma_k$ is one of the
shortest segments connecting $\Gamma_k$ to one of the components
$G_{k+1},G_{k+2},\ldots$, one must have $d(\widetilde{V},V')\leq
d(\widetilde{V},W)$, which entails
\[ i_1+k_2\geq1\,.
\]
This means that the sets $I_1$ and $K_2$ cannot be both empty. If
$K_2\neq\varnothing$, we choose the index 2 in $K_2$, so that the
vertices $V$, $V'$, $W$ and $W'$ constitute a critical quartet. If
$K_2=\varnothing$, $I_1$ is not empty, $\widetilde{V}$ does not
coincide with $V$, and thus $\widetilde{V}$ is a non simple inserted
vertex, necessarily linked in $\Gamma_k$ to a vertex $\widehat{V}$ of
$G$ as displayed in Fig.\,\ref{T}. Choosing now the index 2 in $I_1$,
one finds that the vertices $\widehat{V}$, $V'$, $W$ and $W'$
constitute a critical quartet.

\subsection{Proper $G_c$}

When $G$ is subgraph of a {\sl proper} connected graph $G_c$, the latter
is a tree graph which can be decomposed into the connected components
$G_i$ of $G$ and connecting segments $\Sigma_k$, even when $G_c$ does
not coincide with the specific graph $G_c$ constructed in the previous
section V-A. One needs to distinguish two cases:
\begin{enumerate}[a)]
\item all the segments $\Sigma_k$'s are disjoint. Then all insertions
  are simple and $G_c$ is $G$-simple.
\item at least two segments have one common vertex. This vertex is not
  a simple insertion and thus $G_c$ is not $G$-simple. 
\end{enumerate}

\subsubsection{$G$-simple $G_c$}

Here, as in section IV, we establish part 2.a.i of
Theorem\,\ref{theorem1}  by associating to any chain $\C$ of type $G$ a
particular phase space distribution $\rho_0$ reproducing this
chain. We also give the general form of the $\rho$'s reproducing the chain.

The construction of $\rho_0$ proceeds through the ``Feynman rules'' of
section IV-A, complemented with propagators associated with the segments
of $G_c$. Let $\Sigma$ be such a segment connecting the vertices
$V_\alpha$ and $V_\beta$ of $G$, and let $r$ be its length. Let
$\sigma_\alpha(X,Y)$ and $\sigma_\beta(X,Y')$ be the corresponding
vertex functions of $\C$, where $X$ (resp. $Y$) denote the set of
variables which have the same (resp. a different) assignment in
$\sigma_\alpha$ and $\sigma_\beta$. The compatibility of
$\sigma_\alpha$ and $\sigma_\beta$ allows us to define 
\< \sigma_{\alpha\beta}(X) \equiv \int d^{\kern .1em r}Y\,\sigma_\alpha(X,Y)
   = \int d^{\kern .1em r}Y'\,\sigma_\beta(X,Y')\,. \label{V.3}
\>
For different segments $\Sigma_l$ labelled by the index $l$, we use
the notation $\sigma_{\alpha_l\beta_l}(X_l)$.

To the segment $\Sigma_l$ we now associate the propagator
\< \varpi_l(X_l) = \left\{\begin{array}{ll}
   \displaystyle{1\over\sigma_{\alpha_l\beta_l}(X_l)} &
    \qquad\mbox{if  } X_l\in\S_{\alpha_l\beta_l}\,, \\
   0 & \qquad\mbox{otherwise}\,, \end{array} \right. \label{V.4}
\>
where $\S_{\alpha_l \beta_l}$ is the support of $\sigma_{\alpha_l\beta_l}$. 
This amounts to consider $\Sigma_l$ as a new kind of link (in the
graph $G_c$) which we call {\sl composite link}. Then the function
$\rho_0$ reads
\< \rho_0(\qp) = \left(\prod_{\alpha=1}^n \sigma_\alpha(Z_\alpha)\right)\,
   \left(\prod_{l=1}^{n-1} \varpi_l(X_l)\right)\,\zeta(T')\,, \label{V.5}
\>
where the second product in the r.h.s. is performed on all links,
namely the links of $G$ and the composite links of $G_c$. Note that,
since $G_c$ is a connected tree, the total number of links is $(n-1)$.
As for the function $\zeta(T')$, which is arbitrary but non negative
and normalized, it takes care of the ``passive'' variables
$T=\{x_n,x_{n+1},\ldots,x_N\}$ and their conjugate $T'$, as in
eq.\,\rf{IV.03}.

The proof that the $\rho_0$ of eq.\,\rf{V.5} solves the equations
\rf{II.2} relies on the ``peeling process'' described in section
IV-A. Here, this process has to be extended to the case where the
one-leg vertex $\widehat{V}$ introduced there is attached to a
composite link. Let $\hat{\sigma}(X,Y)$ be the vertex function of $\C$
associated to $\widehat{V}$, where $X$ (resp. $Y$) is the set of
variables whose assignment does not change (resp. changes) through the
composite link. Then eqs.\,\rf{IV.3} become
\< \int d^{\kern .1em r}Y\,\rho_0^{(m)}(\qp) = \rho_0^{(m-1)}(X,X',Y')\,,
 \label{V.6}
\>
and the rest of the proof is completely similar to that given in
section IV-A.

The determination of the general solution $\rho$ of eqs.\,\rf{II.2} is
also carried out along the lines followed in section IV-B, by using
this time the extended peeling process. The definitions of the
measures $d\mu_\alpha$ and of the projectors $P_\alpha$ (which now
involve the function $\rho_0$ of eq.\,\rf{V.5}) are
unchanged. The  Lemma\,\ref{lemma2} still holds, though with an
extended acceptation of ``contiguity'' : two vertices $V_{\alpha_l}$
and $V_{\beta_l}$ of $G$ connected by a composite link $\Sigma_l$ of
$G_c$ are declared contiguous. Actually, only minor changes are needed
to generalize the proof given in Appendix A
(essentially the substitution $x_i\rightarrow Y$). The operator $\Pi$
is now defined as
\< \Pi = 1-
  \sum_{\alpha=1}^n P_\alpha + \sum_{l=1}^{n-1}\,P_{\alpha_l}\,P_{\beta_l}
  \,, \label{V.7}
\>
where the last sum in the r.h.s. is a sum over all links of $G_c$,
composite or not. Its properties \rf{IV.017} and \rf{IV.021} remain
true and, with $\rho$ written as in eq.\,\rf{IV.7}, one finds that the
general solution for $h$ is given again by eq.\,\rf{IV.023}. A change
then occurs in eq.\,\rf{IV.20} when the two
vertices $V_{\alpha_i}$ and $V_{\beta_i}$ there become two vertices
$V_{\alpha_l}$ and $V_{\beta_l}$ connected by a composite link. In this
case eq.\,\rf{IV.20} becomes
\<\int d^{\kern .05em N-r_l}\!X_l'\ \rho_0(X_l,X_l',Y_l,Y_l') =
  {\sigma_{\alpha_l}(X_l,Y_l)\,\sigma_{\beta_l}(X_l,Y_l')\over
  \sigma_{\alpha_l\beta_l}(X_l)}\,. \label{V.8}
\>
One ends up with the following expression of $h$, generalizing the
representation formula \rf{IV.23}
\< \begin{array}{rcl}
  h(\qp) &=&
  f(\qp)-\displaystyle{\sum_{\alpha=1}^n {1\over\sigma_\alpha(Z_\alpha)}}
  \int d^{\kern .05em N}\!Z_\alpha'\,\rho_0(\qp)\,f(\qp) \\
  && +\displaystyle{\sum_{l=1}^{n-1} {1\over\sigma_{\alpha_l\beta_l}(X_l)}}\,
  \int d^{\kern .05em N-r_l}\!X_l'\,d^{\kern .1em r_l}Y_l\,
  d^{\kern .1em r_l}Y_l'\ \rho_0(\qp)\,f(\qp)\,. \end{array} \label{V.9}
\>

We remind the reader that, in this formula:
\begin{enumerate}[i)]
\item f is an arbitrary function in $L^1(\R^{2N},\rho_0\,
  d^{\kern .05em N}\!q\,d^{\kern .05em N}\!p)$;
\item the first sum in the r.h.s. is over all vertices $V_\alpha$ of
  $G$; $Z_\alpha$ denotes the collection of arguments of the vertex
  function $\sigma_\alpha$ and $Z_\alpha'$ the collection of conjugate
  arguments;
\item the second sum is over all links (of length $r_l$), that is the
  simple links   of $G$ and the composite links of $G_c$; the
  definition of the functions $\sigma_{\alpha_l\beta_l}$ occuring in
  the sum, as well as the meaning of the collections of variables
  $X_l$, $X_l'$, $Y_l$ and $Y_l'$, are provided by eq.\,\rf{V.3},
  which reduces to eq.\,\rf{IV.21} in the case of a simple link of index i.
\end{enumerate}

\subsubsection{Non $G$-simple $G_c$}

It remains to prove part 2.a.ii of Theorem\,\ref{theorem1}.

Consider first an arbitrary {\sl quantum} chain $\C$ of type $G$. The
CCS-distributions of $\C$ are then expressed in terms of some density
operator $\hat{\rho}$, in accordance with eqs.\,\rf{I.4}. But, in this
case, a set of CCS-distributions associated
with the insertions of $G_c$ can also be computed through these
equations. One thus obtains an extended chain $\C_c$ of compatible
CCS-distributions associated with all the vertices of $G_c$. Since
$G_c$ is proper, the chain $\C_c$ is admissible (irrespective of the
fact that $G_c$ is not $G$-simple), which implies the admissibility of
$\C$. Hence the quantum admissibility of the graph $G$.

To prove that $G$ is not fully admissible, we show that there are (non
quantum) chains of type $G$ which are not admissible.

\def\rndbox#1#2{\beginpicture \setcoordinatesystem units <.1in,.1in>
   \put {\oval(#1,13.2)} at 0 0 \put {\ninerm #2} at 0 .45 \endpicture}

By assumption $G_c$ contains at least one insertion $V$ with $k\geq3$
legs, say the vertex \rndbox{43}{$\ 12\ldots N\ $} with legs carrying the
indices $j=1,2,\ldots,k$. These legs connect $V$ to $k$ subgraphs
$G_c^{(j)}$ of $G_c$ which are proper connected trees, but mutually
disconnected (otherwise $G_c$ would contain loops). By removing all the
insertions, together with their legs, from each of the $G_c^{(j)}$'s,
we obtain $k$ subgraphs $G^{(j)}$ of $G$ which are proper (not
necessarily connected) trees. The vertices $V_l^{(j)}$ of each
$G^{(j)}$ are of the form \rectbox{2.25}{$\,1,2,\ldots,j-1,j',j+1,\dots,
k,\J_l^{(j)}\,$}\,,\vbox{\hrule height 15pt depth 10pt width 0pt} where
$\J_l^{(j)}$ represents a set $\{k+1,k+2,\ldots,N\}$ of indices, 
each one being primed or not.

We now use the following lemma, the proof of which is given in
\ref{Appendix lemma3}
\begin{lemma}{\ }

In a $2k$-dimensional phase space with $k\geq3$, there exist
$k$-chains of compatible distributions \{$\tau_1(p_1,q_2,q_3,\ldots,q_k)$, 
$\tau_2(q_1,p_2,\-q_3,\ldots,q_k)$, $\ldots$,
$\tau_k(q_1,q_2,\ldots,q_{k-1},p_k)$\} which are not admissible.
\label{lemma3}
\end{lemma}

Let us construct a chain $\C$ of type $G$ by assigning to each vertex
$V_l^{(j)}$ of $G^{(j)}$ ($j=1,\ldots,k$) the CCS-distributions
\< \sigma_l^{(j)}(q_1,\ldots,q_{j-1},p_j,q_{j+1},\ldots,q_k,X_l^{(j)})
  = \tau_j(q_1,\ldots,q_{j-1},p_j,q_{j+1},\ldots,q_k)\,
  \tilde{\sigma}_l^{(j)}(X_l^{(j)})\,, \label{V.10}
\>
where $X_l^{(j)}$ denotes the set of variables $x_i$ corresponding to
$\J_l^{(j)}$ and the $\tilde{\sigma}_l^{(j)}$'s are arbitrary
probability distributions depending on these variables, only subjected
to the apposite compatibility conditions\footnote{That such
$\tilde{\sigma}_l^{(j)}$'s always exist is easy to see, e.g. by
choosing completely factorized forms for them.}. The elements of the
chain $\C=\{\sigma_l^{(j)}\}_{j=1,\ldots,k;l}$ are evidently
compatible CCS-distributions. Let us pretend that these distributions
are marginals of some phase space density $\rho$. Then, by defining
the reduced density
\< \tilde{\rho}(q_1,\dots,q_k,p_1,\ldots,p_k) =
  \int dq_{k+1}dp_{k+1}\ldots dq_Ndp_N\,\rho(\qp) \label{V.11}
\>
in a $2k$-dimensional phase space, one finds that
\begin{eqnarray}
  \tau_j &=& \int d^{N-k}\!X_l^{(j)}\,\sigma_l^{(j)} \qquad ({\rm any\ }l)
  \,, \nonumber \\
  &=& \int dp_1\ldots dp_{j-1}dq_j\,dp_{j+1}\ldots dp_k\,\tilde{\rho}
  \qquad (j=1,\ldots,k)\,. \label{V.12}
\end{eqnarray}
This would mean that the reduced $k$-chain $\widetilde{\C}=
\{\tau_j\}_{j=1,\ldots,k}$ is always admissible, in contradistinction
with Lemma \ref{lemma3}. We conclude that there exist chains of type
$G$ which are not admissible.

The two statements in part 2.a.ii of Theorem\,\ref{theorem1} are now
established and the proof of this theorem is complete.

Finally, we would like to obtain an explicit expression of all
the phase space densities $\rho$ solving eqs.\,\rf{II.2} for a given
quantum chain $\C$ of type $G$, by following again the method of section
IV. However, serious complications crop up in the final step of the
procedure.

First, a particular solution $\rho_0$ is obtained by applying the
formula \rf{IV.03} {\sl to the extended chain} $\C_c$. Alternatively,
one can determine other particular solutions $\rho_0'$ by using,
instead of $\C_c$, the chains $\C_c'$ obtained from $\C_c$ by removing
all or some of the {\sl simple} insertions of $G_c$ and applying the
procedure of section V-B-1 involving composite links and their
associated propagators. Whatever $\rho_0$ is chosen, we keep writing
the general solution in the form $\rho=\rho_0(1+\lambda h)$, as in
eq.\,\rf{IV.7}.

 Then a change appears in the determination of the function $h$,
 because one does not have to require the density $\rho$ to reproduce
 all the CCS-distributions of the chain $\C_c$ (or $\C_c'$), but only
 those of the given chain $\C$. This means that $h$ should satisfy the
 eqs.\,\rf{IV.022}, where the index  $\alpha$ now refers to the only
 elements of the {\sl initial chain} $\C$. As a consequence, the form
 \rf{IV.15} of the appropriate operator $\Pi$ (to be used in
 eq.\,\rf{IV.023}) is no longer valid, since the  properties
 \rf{IV.017} and \rf{IV.021} hold only if the underlying  graph is
 {\sl connected}. Notice that a similar difficulty already appeared
 when dealing with $G$-simple graphs $G_c$ in the previous
 subsection. There, it was overcome by introducing composite links
 which eventually allowed us to remove the insertions of
 $G_c$. Unfortunately, no such device presents itself for non
 $G$-simple $G_c$'s, and constructing the ``good'' projector $\Pi$ in
 this case seems to be quite a difficult problem, which we leave
 unsolved here.

 Of course, the projector $\Pi_c$ associated with the chain $\C_c$
 already provides us with a large class of solutions, but certainly
 not all the solutions.

\section{Conclusions}
\setcounter{equation}{0}

We have investigated the extent to which it is possible to reproduce a
given set of joint probability distributions
$\sigma_\alpha(x_1,x_2,\ldots,x_N)$ with $x_i=q_i$ or $p_i$, in arbitrary
number $n$ and with arbitrary position-momentum assignments of the
$x_i$'s, as marginals of some probability density $\rho(\qp)$ in
$2N$-dimensional phase space. We have been able to give a complete
characterization of those sets which can always be reproduced by a
$\rho(\qp)$ (admissible sets), irrespective of the functional form of the
$\sigma_\alpha$'s provided they are compatible, and both for quantum
probability distributions $\sigma_\alpha$ and for more general
(classical) ones. This has been achieved by introducing a specific,
powerful diagrammatic method and by relying on previous results
\cite{AMRSl}-\cite{AMRS4} obtained in the case $N=2$ by means of
Bell-like inequalities in phase space.

When both classical and quantum sets
$\{\sigma_\alpha\}_{\alpha=1,\ldots,n}$ are admissible, we have
constructed the general solution $\rho(\qp)$ of the problem. When only
quantum sets are admissible, we have the explicit expression of a
large class of solutions, which however is not exhaustive. Concerning
the dynamical aspect which is completely ignored in this paper, our
results in the quantum case motivate the construction of realistic
quantum mechanics reproducing $(N+1)$ marginals at all times $t$ and
thus considerably improving on the de~Broglie-Bohm mechanics
\cite{dBB}, which reproduces only one $\sigma_\alpha$ (the position
probability distribution $\sigma(\overrightarrow{\kern -.1em 
q\kern .1em},t)$).

On the other hand, all cases of non admissibility have been
identified. For quantum $\sigma_\alpha$'s again, this may be viewed as
a general contextuality theorem of the Gleason-Kochen-Specker type
\cite{GKS}, which also extends a previous result of this type due to
Martin and Roy \cite{MR}. At the same time, this provides a proof of a
long-standing conjecture, the ``($N+1$) marginal theorem''.

From the mathematical standpoint, the parts of our main theorem
(Theorem\,\ref{theorem1}) pertaining to the quantum case are
essentially new statements concerning multi-dimensional Fourier
transforms in $L^2(\R^{2N},d^{\kern .05em N}\!q\,d^{\kern .05em
N}\!p)$. These statements vastly extend the results of Cohen and
Zaparovanny \cite{CZ} for two non-intersecting marginals to the case of
$N+1$ marginals containing overlapping variables. Thus, they can be
expected to  open new applications in classical signal and image
processing \cite{C}. From
the physical point of view, our results completely settle, at a formal
level, the question of ``maximal reality'' raised and already
investigated in special cases \cite{RS1}-\cite{AMRS4}. Their possible
relevance for related fundamental problems of quantum theory (in
particular for helping towards a clarification of the still
controversial problem of measurement) remains to be explored.

\newpage\appendix
\def\thesection{Appendix \Alph{section}.}
\def\thesubsection{\Alph{section}. \alph{subsection})}
\def\theequation{\Alph{section}.\arabic{equation}}

\section{Proof of Lemma \ref{lemma2}}
\label{Appendix lemma2}
\setcounter{equation}{0}

\subsection{Commutation relation \rf{IV.014}}
Consider two contiguous vertices $V_\alpha$ and $V_\beta$ of $G$
connected by a link $l_i$ with index $i$, and denote by
$\sigma_\alpha(x_i,X,T)$ and $\sigma_\beta(x'_i,X,T)$ the
corresponding distributions of the chain $\C_n$, where
$X=\{x_1,\ldots,x_{i-1},x_{i+1},\ldots,x_{n-1}\}$. By removing the
link $l_i$, the proper tree graph $G$ is broken in two connected
components $G_\alpha$ and $G_\beta$ : $G=G_\alpha\cup l_i\cup G_\beta$.
To this splitting clearly corresponds a partition $\{X_\alpha,X_\beta\}$
of the variables $X$ such that, among $X$ and their conjugate $X'$,
all the variables $\{X_\alpha,X'_\alpha,X_\beta\}$ and only them
appear in the vertices of $G_\alpha$, whereas all the variables
$\{X_\alpha,X_\beta,X'_\beta\}$ and only them appear in the vertices
of $G_\beta$. In parallel, the particular solution $\rho_0$ given in
eq. \rf{IV.03} factorizes as
\<
  \rho_0 = {\rho_\alpha\,\rho_\beta \over\sigma_{\alpha\beta}}\,\zeta\,,
  \label{A.1}
\>
where the function $\rho_\alpha$ (the ``$\rho_0/\zeta$'' of the subchain
of $\C_n$ of type $G_\alpha$) depends only on
($x_i,X_\alpha,X'_\alpha,X_\beta$), the function $\rho_\beta$ depends
only on ($x'_i,X_\alpha,X_\beta,X'_\beta$), and
\[
  \sigma_{\alpha\beta}(X,T) = \int dx_i\,\sigma_\alpha(x_i,X,T)
  = \int dx'_i\,\sigma_\beta(x'_i,X,T)\,.
\]
The measures $d\mu_\alpha$ and $d\mu_\beta$ as defined by
eq. \rf{IV.6} now take the form :
\<
  \left\{ \begin{array}{rcl}
  d\mu_\alpha &=& {\displaystyle\rho_\alpha\over\displaystyle\sigma_\alpha}
  \,dX'_\alpha\,{\displaystyle1\over\displaystyle\sigma_{\alpha\beta}}
  \,\rho_\beta\,dX'_\beta\,dx'_i\,\zeta\,dT'\,, \\
  d\mu_\beta &=& \rho_\alpha\,dX'_\alpha\,
  {\displaystyle1\over\displaystyle\sigma_{\alpha\beta}}\,
  {\displaystyle\rho_\beta\over\displaystyle\sigma_\beta}\,
  dX'_\beta\,dx_i\,\zeta\,dT'\,.
  \end{array} \right. \label{A.2}
\>
Hence, for any $g(\qp)\in L^1(\R^{2N},\rho_0\,d^{\kern .05em
  N}\!q\,d^{\kern .05em N}\!p)$ :
\[
  P_\alpha P_\beta\,g =
  {\displaystyle 1\over\displaystyle\sigma_{\alpha\beta}}\,\int dx'_i\,
  \int dX'_\alpha\,{\displaystyle\rho_\alpha\over\displaystyle\sigma_\alpha}
  \int dX'_\beta\,\rho_\beta\,\int dT'\,\zeta\,(P_\beta\,g)\,.
\]
The integrations over $X'_\alpha$, $X'_\beta$ and $T'$ can be
performed explicitly since $P_\beta\,g$ does not depend on these
variables. Noticing that
\<
  \int dX'_\alpha\,\rho_\alpha = \sigma_\alpha\,, \qquad\qquad
  \int dX'_\beta\,\rho_\beta = \sigma_\beta\,, \label{A.3}
\>
and taking account of eq.\,\rf{IV.04}, we get
\begin{eqnarray}
  P_\alpha P_\beta\,g &=&
  {\displaystyle 1\over\displaystyle\sigma_{\alpha\beta}}\,
  \int dx'_i\,\sigma_\beta\,(P_\beta\,g)\,, \label{A.4} \\
  &=& {\displaystyle 1\over\displaystyle\sigma_{\alpha\beta}}\,
  \int dx'_i\,\sigma_\beta\,
  {\displaystyle 1\over\displaystyle\sigma_{\alpha\beta}}\,
  \int dx_i\,dX'_\alpha\,dX'_\beta\,dT'\,\rho_\alpha
  {\displaystyle \rho_\beta\over\displaystyle\sigma_\beta}\,
  \zeta\,g\,. \label{A.5}
\end{eqnarray}
Since $\sigma_\beta$ does not depend on $x_i$, $X'_\alpha$ and
$X'_\beta$, the factors $\sigma_\beta$ and $1/\sigma_\beta$ in
eq.\,\rf{A.5} cancel each other. This gives
\<
  P_\alpha P_\beta\,g =
  {\displaystyle 1\over\left(\displaystyle\sigma_{\alpha\beta}\right)^2}\,
  \int dx_i\,dx'_i\,dX'_\alpha\,dX'_\beta\,dT'\,\rho_\alpha\,\rho_\beta\,
  \zeta\,g\,. \label{A.6}
\>
The r.h.s. of this equation is symmetric in $\alpha\leftrightarrow\beta$,
which establishes eq.\,\rf{IV.014}.

\subsection{Relation \rf{IV.015}}
Let $V_\alpha$, $V_\beta$ and $V_\gamma$ be now three vertices of $G$
such that $V_\alpha$ belongs to the path connecting $V_\gamma$ to
$V_\beta$ and is contiguous to $V_\beta$. Consider again the connected
subgraphs $G_\alpha$ and $G_\beta$ defined in A.a) above, together
with the partition $\{X_\alpha,X_\beta\}$ of the variables $X$, and
distinguish in $G_\alpha$ the linear subgraph $G_{\alpha\gamma}$ made
of the vertices $V_\alpha$, $V_\gamma$ and the path connecting
them. Denote by $I_{\alpha_1}$ the set of indices of the links of
$G_{\alpha\gamma}$ and by $I_{\alpha_2}$ the set of indices of the
remaining links in $G_\alpha$. To this splitting corresponds a further
partition $\{X_{\alpha_1},X_{\alpha_2}\}$ of the variables $X_\alpha$,
as indicated in Fig. \ref{FigApp} :

\def\rarrow
{\beginpicture
   \setlinear \plot -0.1 0  -1 .3  -.8 0  -1 -.3 -0.1 0 /
   \setdots <2pt>
   \putrule from 0 .1 to 0 1.6
\endpicture}

\def\larrow
{\beginpicture
   \setlinear \plot 0.1 0  1 .3  .8 0  1 -.3 0.1 0 /
   \setdots <2pt>
   \putrule from 0 .1 to 0 1.6
\endpicture}

\def\figAppendix{
\beginpicture
\setcoordinatesystem units <.1in,.1in>
\put {\rectbox{1}{$\scriptstyle x_i,X'_{\alpha_1},X_{\alpha_2},X_\beta,T$}}
 [rb] at 0 -0.3
\put {\rectbox{1}{$\scriptstyle x_i,X_{\alpha_1},X_{\alpha_2},X_\beta,T$}}
 [rb] at 14.5 -0.3
\put {\rectbox{1}{$\scriptstyle x'_i,X_{\alpha_1},X_{\alpha_2},X_\beta,T$}}
 [lb] at 17 -0.3
\putrule from 14.5 0 to 17 0
\put {\larrow} at -10.5 -1.945
\put {\rarrow} at 14.5 -1.945
\put {\larrow} at 14.5 -1.945
\put {\rarrow} at 17 -1.945
\putrule from -9.6 -2.8 to 13.6 -2.8
\putrule from 15.4 -2.8 to 16.1 -2.8
\setdashes <1.52pt>
\putrule from 0 0 to 4 0
\put {$\scriptstyle I_{\alpha_1}$} [b] at 2 .3
\put {$\scriptstyle i$} [b] at 15.75 .3
\put {$\scriptstyle V_\gamma$} [b] at -5.2 1.7
\put {$\scriptstyle V_\alpha$} [b] at 9.2 1.75
\put {$\scriptstyle V_\beta$} [b] at 22.25 1.6
\put {$\scriptstyle G_{\alpha\gamma}$} at 2 -3.9
\put {$\scriptstyle l_i$} at 15.8 -3.9
\endpicture}

\begin{figure}[h]
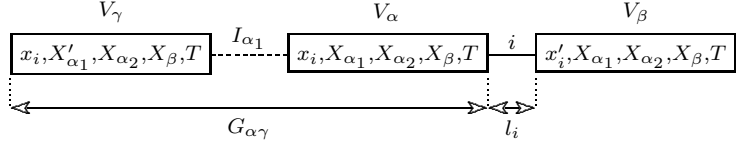

\centerline{\figAppendix}
\caption{The path between $V_\gamma$ and $V_\beta$ in $G$. \label{FigApp}}
\end{figure}

With a factorization of the measure $d\mu_\gamma$ analogous to those
of eq.\,\rf{A.2}, we can write
\<
  P_\gamma P_\alpha P_\beta\,g =
  \int dX_{\alpha_1}\,dX'_{\alpha_2}\,dX'_\beta\,dx'_i\,dT'\,
  {\displaystyle\rho_\alpha\over\displaystyle\sigma_\gamma}\,
  {\displaystyle 1\over\displaystyle\sigma_{\alpha\beta}}\,
  \rho_\beta\,\zeta\,(P_\alpha P_\beta\,g)\,. \label{A.7}
\>
Here, we can perform explicitly the integrations over $x'_i$,
$X'_\beta$, $X'_{\alpha_2}$ and $T'$, for $P_\alpha P_\beta\,g$ does
not depend on these variables. First :
\<
  \int dx'_i\,dX'_\beta\,\rho_\beta =
  \int dx'_i\,\sigma_\beta = \sigma_{\alpha\beta}\,. \label{A.8}
\>
The left equality in eq.\,\rf{A.8} results, as in eq.\,\rf{A.3}, from
the ``peeling process'' (described in section IV.A) corresponding to
the reduction of the graph $G_\beta$ to the vertex
$V_\beta$. Similarly, the (partial) peeling process corresponding to
the reduction $G_\alpha\rightarrow G_{\alpha\gamma}$ yields
\<
  \int dX'_{\alpha_2}\,\rho_\alpha = \rho_{\alpha\gamma}\,, \label{A.9}
\>
where $\rho_{\alpha\gamma}$ is the ``$\rho_0/\zeta$'' of the subchain
of type $G_{\alpha\gamma}$. Thanks to eqs.\,\rf{A.8}, \rf{A.9} and
\rf{IV.4}, equation \rf{A.7} boils down to
\[
  P_\gamma P_\alpha P_\beta\,g = \int dX_{\alpha_1}\,
  {\displaystyle\rho_{\alpha\gamma}\over\displaystyle\sigma_\gamma}\,
  (P_\alpha\,P_\beta\,g)\,
\]
or, by inserting the expression \rf{A.4} of $P_\alpha P_\beta g$ :
\<
  P_\gamma P_\alpha P_\beta\,g = \int dX_{\alpha_1}\,
  {\displaystyle\rho_{\alpha\gamma}\over\displaystyle\sigma_\gamma}\,
  \int dx'_i\,
  {\displaystyle\sigma_\beta\over\displaystyle\sigma_{\alpha\beta}}\,
  (P_\beta\,g)\,. \label{A.10}
\>
On the other hand :
\<
  P_\gamma P_\beta\,g =
  \int dX_{\alpha_1}\,dX'_{\alpha_2}\,dX'_\beta\,dx'_i\,dT'\,
  {\displaystyle\rho_0\over\displaystyle\sigma_\gamma}\,
  \zeta\,(P_\beta\,g)\,, \label{A.11}
\>where the integrations over $X'_\beta$, $X'_{\alpha_2}$ and $T'$ can
be performed explicitly since $P_\beta\,g$ does not depend on these
variables. The integration over $X'_\beta$ first produces, through the
partial peeling process corresponding to $G\rightarrow(G_\alpha\cup
l_i\cup V_\beta)$ :
\<
  \int dX'_\beta\,\rho_0 = \rho_\alpha\,
  {\displaystyle 1\over\displaystyle\sigma_{\alpha\beta}}\,
  \sigma_\beta\,. \label{A.12}
\>
Then eqs.\,\rf{A.9} and \rf{IV.04} are used again for the integrations
over $X'_{\alpha_2}$ and $T'$ respectively. Altogether, this reduces
the expression \rf{A.11} to the r.h.s. of eq.\,\rf{A.10}. Therefore
$P_\gamma P_\alpha P_\beta\,g = P_\gamma P_\beta\,g$, which
establishes eq.\,\rf{IV.015} in the case where $V_\alpha$ is
contiguous to $V_\beta$.

The proof of eq.\,\rf{IV.015} in the case where $V_\alpha$ is contiguous
to $V_\gamma$ is completely similar.

\hfill q.e.d.

\section{Proof of Lemma \ref{lemma3}}
\label{Appendix lemma3}
\setcounter{equation}{0}

We construct a particular $k$-chain of compatible distributions $\tau_j$ and
we prove that there is no positive phase space density $\rho$
reproducing these distributions as marginals. We take $\tau_j$ of the form:
\[ \tau_j(q_1,\ldots,q_{j-1},p_j,q_{j+1},\ldots,q_k) = 
   \gamma_j(p_j)\,\taub_j(q_1,\ldots,q_{j-1},q_{j+1},\ldots,q_k)
   \quad(j=1,\ldots,k)
\]
where $\{\taub_1,\ldots,\taub_k\}$ is a $k$-chain of compatible, reduced 
distributions and the $\gamma_j(p_j)$'s are arbitrary, normalized one variable
distributions.

Let us look for a phase space density $\rho$ reproducing the $\tau_j$'s.
The $\taub_j$'s are given in terms of the configuration space density
\< \rhob(q_1,\ldots,q_k)\equiv\int d^kp\,\rho(q_1,\ldots,q_k,p_1,\ldots,p_k)
  \label{L.1}
\>
as
\< \taub_j(q_1,\ldots,q_{j-1},q_{j+1},\ldots,q_k) =
   \int dq_j\,\rhob(q_1,\ldots,q_k)\,. \label{L.2}
\>
We now choose the $\taub_j$'s as follows:
\setbox10=\hbox{$\scriptstyle r=2 $}
\setbox11=\hbox{$\renewcommand{\arraystretch}{.5}
  \begin{array}{cc} \scriptstyle r=1 \\ \scriptstyle r\neq j \end{array}
  \renewcommand{\arraystretch}{1.4}$}
\setbox12=\hbox{$\renewcommand{\arraystretch}{.5}
  \begin{array}{cc} \scriptstyle r=1 \\ \scriptstyle r\neq i,j \end{array}
  \renewcommand{\arraystretch}{1.4}$}
\def\prodrk{\displaystyle\prod_{\copy10}^{k}}
\def\prodrjk{\kern -5pt\displaystyle\prod_{\copy11}^{k}\kern -5pt}
\def\prodrijk{\displaystyle\prod_{\copy12}^{k}\kern -5pt}
\< \left\{\begin{array}{rcl}
   \taub_1(q_2,\ldots,q_k) &=& \prodrk\,T_r-\prodrk\,U_r\,, \\
   \taub_j(q_1,\ldots,q_{j-1},q_{j+1},\ldots,q_k) &=&
     \prodrjk\,T_r + \prodrjk\,U_r\,, \qquad (j=2,\ldots,k)\,,
   \end{array} \right. \label{L.3}
\>

where
\[ \left\{\begin{array}{rll}
   T_r &= &{1\over 2}\left[\delta(q_r-1)+\delta(q_r+1)\right]\,, \\
   U_r &= &{1\over 2}\left[\delta(q_r-1)-\delta(q_r+1)\right]\,,
   \end{array} \right. \quad (r=1,\ldots,k)\,.
\]
The $\taub_j$'s, which appear as sums of $2^{k-2}$ monomials of the form
\[ {1\over 2^{k-2}}\,\prodrjk \delta(q_r-\varepsilon_r)
   \qquad \varepsilon_r=\pm 1\,,
\]
are obviously positive and normalized. Furthermore:
\[ \int dq_i\,\taub_j(q_1,\ldots,q_{j-1},q_{j+1},\ldots,q_k) =
   \prodrijk T_r \qquad (i\neq j)\,.
\]
The r.h.s. of this equation is symmetric in $i\leftrightarrow j$, which
entails the compatibility of the $\taub_j$'s.

Clearly, the most general {\sl positive} $\rhob$ obeying equations
\rf{L.2} is the sum of $2^k$ terms proportional to
$\prod_{r=1}^k\delta(q_r-\varepsilon_r)$. Equivalently, $\rhob$ can be
written as a homogeneous polynomial 
$P(\{T_r\},\{U_r\})$ of degree $k$ which, for each index $r$, is
linear in $T_r$ and $U_r$. Then, since $\int\!dq_j T_j=1$ and
$\int\!dq_j U_j=0$, we can express $\int\!dq_j\,\rhob$ as $\partial
P/\partial T_j$, so that eqs.\,\rf{L.2} and \rf{L.3} yield:
\[ \left\{\begin{array}{rll}
   \displaystyle{\partial P\over\partial T_1\mathstrut} &=
   &\prodrk\,T_r-\prodrk\,U_r\,, \\
   \displaystyle{\partial P\over\partial T_j\mathstrut} &= 
   &\prodrjk\,T_r+\prodrjk\,U_r\,, \qquad (j=2,\ldots,k).
   \end{array}\right.
\]
The general solution of these equations is:
\< P=\prod_{r=1}^k\,T_r-T_1\prodrk\,U_r+\sum_{j=2}^k T_j
   \prodrjk\,U_r +\lambda\prod_{r=1}^k U_r\,, \label{L.4}
\>
where $\lambda$ is an arbitrary real parameter.

Now, whatever the value of $\lambda$ is, $P$, and thus $\rho$, are not
positive. To show this, it is sufficient to look at the coefficients
of the two monomials
\[ \delta(q_1+1)\prod_{r=2}^k\delta(q_r-1) \qquad \hbox{and} \qquad 
   \delta(q_1-1)\,\delta(q_2+1)\,\delta(q_3+1)\prod_{r=4}^k\delta(q_r-1)
\]
which appear in eq.\,\rf{L.4} if $k\geq3$. One finds
$-(k-1+\lambda)/2^k$ and $(k-5+\lambda)/2^k$ respectively, the sum of
which is independent of $\lambda$ and negative.

\hfill q.e.d.

\end{document}